\documentclass[aps,prb,twocolumn,superscriptaddress]{revtex4-2}

\usepackage[hidelinks]{hyperref}
\hypersetup{
	colorlinks,
	linkcolor={red},
	citecolor={blue},
	urlcolor={blue}
}
\usepackage{footnote}
\usepackage{physics}
\usepackage{balance}
\usepackage{amssymb}
\usepackage{suffix}
\usepackage{mathtools}
\usepackage[utf8]{inputenc}
\usepackage{booktabs}
\usepackage{cases}
\usepackage[multiple]{footmisc}
\usepackage{dcolumn}
\usepackage{color,soul}
\usepackage{rotating}
\usepackage{perpage}
\usepackage{siunitx}
\usepackage{xcolor}
\usepackage{soul}
\usepackage{amsmath}
\usepackage[T1]{fontenc}
\usepackage{etoolbox}
\usepackage{graphics}
\usepackage{siunitx}
\usepackage{float}	
\usepackage{collref}
\usepackage{multirow}
\usepackage{mathtools}
\usepackage{bm}
\usepackage{url}

\usepackage{tikz}
\usepackage{tikz-3dplot}
\usepackage{accents}

\usepackage{tikz}
\usetikzlibrary{arrows.meta, positioning, decorations.pathmorphing, shapes.geometric}
\makeatletter
\newcommand{\doublewidetilde}[1]{{%
		\mathpalette\double@widetilde{#1}%
}}
\newcommand{\double@widetilde}[2]{%
	\sbox\z@{$\m@th#1\widetilde{#2}$}%
	\ht\z@=.9\ht\z@
	\widetilde{\box\z@}%
}
\makeatother

\newcommand{\MY}[1]{\textcolor{green!60!black}{\fbox{\textbf{Mohsen}} {\sl#1}}}
\newcommand{\JD}[1]{\textcolor{blue}{\fbox{Jakob} {\sl#1}}}

\usepackage{etoolbox,lipsum}

\newcommand{\red}[1]{\textcolor{red}{#1}}

\newcommand{\ddt}{\frac{d}{dt}}

\usepackage{braket}
\usepackage{physics}
\newcommand{\bk}{k}
\newcommand{\bp}{p}
\newcommand{\bq}{q}
\newcommand{\bl}{l}

\newcommand{\qir}{X_\text{IR}}
\newcommand{\qr}{X_\text{R}}
\newcommand{\pir}{P_\text{IR}}
\newcommand{\pr}{P_\text{R}}

\newcommand{\bqir}{\bar{X}_\text{IR}}
\newcommand{\bqr}{\bar{X}_\text{R}}
\newcommand{\bpir}{\bar{P}_\text{IR}}
\newcommand{\bpr}{\bar{P}_\text{R}}

\newcommand{\qirq}{X_{\text{IR},\bq}}
\newcommand{\qrq}{X_{\text{R},\bq}}
\newcommand{\pirq}{P_{\text{IR},\bq}}
\newcommand{\prq}{P_{\text{R},\bq}}

\newcommand{\Oa}{\text{A}}
\newcommand{\Ob}{\text{B}}
\newcommand{\Oc}{\text{C}}
\newcommand{\R}{\text{R}}
\newcommand{\IR}{\text{IR}}
\newcommand{\s}{\text{s}}

\newcommand{\Bk}{B_\bk}
\newcommand{\Ak}{A_\bk}

\definecolor{dgreen}{RGB}{0,190,0} 

\begin{document}
	\title{Anomalous dynamical energy flows via nonlinear phononics in spin-Peierls chains}
	
	\author{Jakob Dolgner}
	\email{j.dolgner@fkf.mpg.de}
	\address{Max Planck Institute for Solid State Research, D-70569, Stuttgart, Germany}
	
	\author{Dirk Manske}
	\address{Max Planck Institute for Solid State Research, D-70569, Stuttgart, Germany}
	
	\author{James K. Freericks}
	\address{Department of Physics, Georgetown University, Washington DC 20057, USA}
	
	\author{Mohsen Yarmohammadi}
	\email{mohsen.yarmohammadi@georgetown.edu}
	\address{Department of Physics, Georgetown University, Washington DC 20057, USA}
	\date{\today}
	
	\begin{abstract}
		We investigate the nonequilibrium spin-phonon dynamics and energy cascades in a strongly dimerized spin-Peierls chain using a multi-tiered nonlinear phononics architecture. To bypass linear selection rules prohibiting the direct excitation of the Raman-active dimerization mode, a terahertz laser drives an infrared mode that nonlinearly couples to the Raman lattice displacement, subsequently modulating the magnetic exchange. Employing a bond-operator formalism with a second-order cumulant expansion of the Lindblad master equation, we show that maximum energy transfer into the magnetic sector is governed by a dynamical impedance-matching condition rather than the unperturbed triplon density of states. We find that the sustained energy input of continuous-wave driving builds high excitation densities that severely back-act on the lattice, overdamping the primary phonon and smearing magnetic features via power broadening. Conversely, the small time-integrated energy of a pulse keeps the response perturbative, terminating before back-action accumulates and preserving sharp Fano-like quantum interferences. These insights establish limits for controlling dynamic magnetic states without quenching the driving lattice modes.
	\end{abstract}
	
	\maketitle
	{\allowdisplaybreaks	
		
		\section{Introduction}
		Laser excitation provides a route to dynamically control strongly correlated systems, accessing macroscopic phases distinct from equilibrium ground states~\cite{Buzzi2018,delaTorre2021,Afanasievetal2021,Mitrano2024,Roelcke2024,Xu2025,doi:10.1126/science.1197294,Fava2024}. Because photons couple weakly to localized magnetic moments in insulators, magnetophononics circumvents this barrier by using lattice vibrations as a proxy for magnetic manipulation~\cite{Afanasievetal2021,PhysRevMaterials.2.064401,PhysRevB.107.184440,PhysRevB.103.045132,10.1063/1.4958846,doi:10.1126/science.adi9601,10.1093/pnasnexus/pgaf002,PhysRevB.107.174415,PhysRevB.110.064420,PhysRevB.110.134442,4ddn-y88c,c3tb-h5hv,doi:10.1126/sciadv.ado0722}. Driving an infrared~(IR)-active dipole resonance with terahertz~(THz) radiation induces atomic displacements that perturb underlying superexchange pathways via spin-phonon coupling~(SPC).
		
		Isolating these feedback mechanisms requires a system with exceptionally strong magnetoelastic coupling. Quasi-one-dimensional~(1D) spin-Peierls chains, such as CuGeO$_3$~\cite{PhysRevB.63.094401,Chen2021,park2025weaklyinteractingspinonstightly,VANLOOSDRECHT19971017,BUCHNER1999956,YUASA20081087} and TiOCl~\cite{PhysRevLett.95.097203,PhysRevB.71.100405,PhysRevLett.107.107402} offer precisely such properties. Below a critical temperature, magnetoelastic forces induce a spatial symmetry breaking in these compounds, generating an alternating bond configuration that locks localized moments into a gapped singlet phase. Continuous optical driving of this broken-symmetry state stabilizes a nonequilibrium steady state (NESS)~\cite{PhysRevB.103.045132}, triggers high-order frequency upconversion~\cite{PhysRevB.110.064420}, and hosts first-order phase transitions~\cite{c3tb-h5hv}. To reach a NESS, the lattice must couple to environmental heat sinks to balance the continuous driving and prevent runaway heating~\cite{PhysRevB.103.045132, Ikeda2020,4ddn-y88c,RevModPhys.97.025004}. Under strong-coupling conditions, the resulting dynamical back-action between the lattice and spin sectors gives rise to composite quasiparticles---hybridized phonon-bitriplons~\cite{PhysRevB.107.174415}. 
		
		Linear selection rules often prevent the direct optical activation of targeted lattice distortions. In centrosymmetric materials, the structural dimerization that opens the spin-Peierls gap is driven entirely by gerade, $A_{1g}$ Raman-active modes in the reduced symmetry group of the dimerized state~\cite{PhysRevB.59.14356,jz36-8kz9,PhysRevB.50.16468,PhysRevB.52.4185}. Because they lack a net electric dipole moment, these modes cannot be excited directly by a laser field. Nonlinear phononics circumvents this limitation~\cite{Henstridge2022,PhysRevResearch.3.L032046,PhysRevB.89.220301,PhysRevLett.118.054101,Rini2007,disa_photoinduced_2023,Basini2024}: resonantly pumping a symmetry-allowed infrared-active mode engages a cubic anharmonic coupling of the form $X_{\mathrm{IR}}^2 X_{\mathrm{R}}$ (where $X$ denotes lattice displacement), exerting both a rectified DC force and a second-harmonic driving force on the Raman-active dimerization coordinate. 
		
		While earlier models by some of us analyzed direct energy transfer from a driven IR phonon to the spin network in a spin-Peierls chain~\cite{PhysRevB.103.045132}, the requirement of an intermediate Raman mode with significantly stronger SPC raises a fundamental question: how does this multi-tiered, nonlinear lattice coupling alter the dynamics of energy routing? This architecture introduces a cascaded pathway in which power injected into the IR mode is nonlinearly transferred to the Raman mode, subsequently depositing into the triplon sector via spin-phonon modulation of the exchange interactions. Resolving the microscopic dynamics of this transient energy cascade is necessary to establish limits on lattice-spin relaxation, optimize magnetic switching efficiency, and suppress parasitic heating that degrades the target phase.\begin{figure*}
			\centering
			\includegraphics[width=0.9\linewidth]{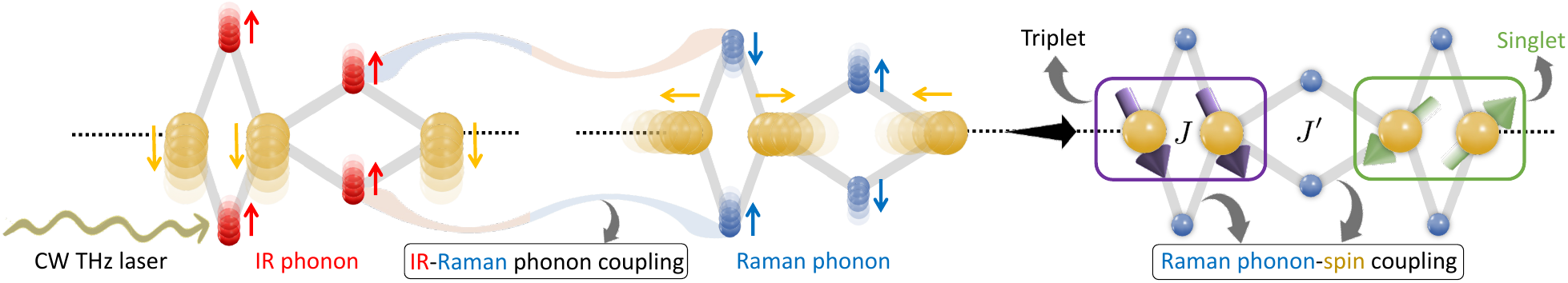}
			\caption{Schematic of the THz laser-driven 1D spin-Peierls system.
				The multi-stage dynamical process is initiated by a continuous-wave (CW) THz laser that resonantly drives a macroscopic, polar IR-active phonon mode (left). Through a symmetry-allowed nonlinear cubic anharmonicity, energy is subsequently transferred from the driven IR mode to a Raman-active $A_{1g}$ dimerization mode (middle). The resulting coherent Raman distortion dynamically modulates the distances between adjacent magnetic ions along the chain (right). This Raman phonon-spin coupling continuously renormalizes the intradimer ($J$) and interdimer ($J'$) exchange interactions, thereby driving the system out of its non-magnetic singlet ground state and generating gapped triplet magnetic excitations.}
			\label{fig:schematic}
		\end{figure*}
		
		In this work, we answer this question by investigating the non-equilibrium spin-phonon dynamics of a strongly dimerized $S=1/2$ spin-Peierls chain under both continuous-wave (CW) and transient pulsed optical excitation. Bypassing the finite-size constraints of exact diagonalization, we model the energy cascade using a bond-operator formalism combined with a second-order cumulant expansion of the Lindblad master equation. Our results demonstrate that, unlike in the previously studied one-phonon system~\cite{PhysRevB.103.045132}, peak energy transfer to the magnetic network is not strictly dictated by the triplon density of states~(DOS) at the bitriplon band edges; rather, it is governed by a dynamical impedance-matching condition between the lattice and spin sectors. By mapping this parameter space across both temporal profiles, we reveal a fundamental split in their physical behavior: sustained CW driving forces the system into a back-action-dominated regime where continuous triplon accumulation overdamps the primary lattice mode and washes out continuum features via power broadening, whereas short pulses restrict energy delivery to keep the response perturbative, terminating before back-action builds and thereby preserving sharp Fano-like quantum interferences. Identifying these distinct operational regimes provides critical guidance for designing high-intensity optical protocols that dynamically control magnetic states without quenching the driving lattice modes.
		
		\section{Theory}\label{s2}
		In quasi-1D spin-Peierls chains, the strong magnetoelastic feedback drives a spatial symmetry breaking below a material-specific critical temperature, locking localized $S=1/2$ moments into a gapped singlet ground state. We benchmark our theoretical framework using two prototypical spin-Peierls systems featuring distinct microscopic exchange mechanisms: Indirect superexchange in $\text{CuGeO}_3$ and direct exchange in $\text{TiOCl}$. The inorganic cuprate $\text{CuGeO}_3$ forms $c$-axis chains of $S=1/2$ $\text{Cu}^{2+}$ ions, where the primary antiferromagnetic exchange $J$ is mediated indirectly through intermediate $\text{O}\ 2p$ orbitals~\cite{PhysRevLett.70.3651,PhysRevB.54.1105}. The extreme sensitivity of $J$ to the $\text{Cu–O–Cu}$ bond geometry facilitates a spin-Peierls transition at a critical temperature of $ 14\,\mathrm{K}$. Conversely, the strongly correlated Mott insulator $\text{TiOCl}$ exhibits a spin-Peierls transition at a significantly higher critical temperature of $ 67\,\mathrm{K}$~\cite{PhysRevB.81.125133,PhysRevLett.95.097203}. In contrast to $\text{CuGeO}_3$, quasi-1D $S=1/2$ chains in $\text{TiOCl}$ run along the $b$-axis, where magnetic interactions are governed by direct $d\text{--}d$ exchange between adjacent $\text{Ti}^{3+}$ ions~\cite{PhysRevB.71.100405}.
		
		To capture the lattice dynamics, we model the participating optical phonons as dispersionless Einstein modes with local, real-space interactions. This approximation is motivated by the kinematics of THz optical pumping. Because the momentum of a THz photon ($\sim 10^4 \text{ m}^{-1}$) is vanishingly small relative to the crystal's Brillouin zone boundaries ($\sim 10^9 \text{ m}^{-1}$), momentum conservation strictly confines direct optical excitations to the macroscopic $q \approx 0$ ($\Gamma$-point) limit. 
		
		\subsection{Hamiltonian model}
		
		We begin by defining the free Hamiltonian for the IR-active phonon branch~[Fig.~\ref{fig:schematic}, left]
		\begin{equation}
			H_{\rm IR} = \frac{1}{2}\sum_\bq \left( \frac{\pirq^2}{m_\IR} + m_\IR \omega_\IR^2\qirq^2 \right)\,,
		\end{equation}
		where the amplitude and canonical momentum variables are defined in terms of bosonic ladder operators (setting $\hbar = 1$) as $X_{\rm IR,\bq} = \frac{1}{\sqrt{2m_{\rm IR}\omega_{\rm IR}}}(b^\dagger_{\rm IR,-\bq} + b_{\rm IR,\bq})$ and $P_{\rm IR,\bq} = i\sqrt{\frac{ m_{\rm IR} \omega_{\rm IR}}{2}}(b^\dagger_{\rm IR,-\bq} - b_{\rm IR,\bq})\,.$ For notational brevity regarding the macroscopically excited IR mode, we omit the $\bq=0$ index for uniform zone-center operators (i.e., $\qir \equiv X_{\IR,\bq=0}$).
		
		The incident laser field $E(t)$ couples linearly and exclusively to this macroscopic IR-active mode~[Fig.~\ref{fig:schematic}, left]. To comprehensively capture both driven steady-state behavior and the transient non-equilibrium dynamics, we consider $E(t)$ as either a CW drive or a finite-duration THz pulse. This direct dipole interaction is given by\begin{equation}\label{eq_2}
			H_{\rm l-IR} = -\sqrt{N} E(t) \qir\,,
		\end{equation}where the effective dipole matrix element is absorbed into $E(t)$. Here, $N$ denotes the number of dimers and the $\sqrt{N}$ prefactor arises from the Fourier transform of the extensive real-space interaction. 
		
		Next, we consider the target Raman modes~[Fig.~\ref{fig:schematic}, middle]. Upon unit cell doubling, the zone-edge phonons driving the lattice deformation fold back to the Brillouin zone center. Recent density functional theory calculations for CuGeO$_3$ identify three principal dimerization modes at 6.05, 10.56, and 21.83~THz~\cite{jz36-8kz9}. Their free dynamics are captured by an analogous Einstein Hamiltonian
		\begin{equation}
			H_{\rm R} = \frac{1}{2}\sum_\bq \left( \frac{\prq^2}{m_\R} + m_\R \omega_\R^2\qrq^2 \right)\,,
		\end{equation}
		with $X_{\rm R,\bq}$ and $P_{\rm R,\bq}$ defined in the same manner as the IR operators.
		
		As gerade modes, these principal dimerization modes are Raman-active but IR-inactive, prohibiting direct optical excitation. This limitation is circumvented via nonlinear phononics~\cite{Henstridge2022,PhysRevResearch.3.L032046,PhysRevB.89.220301,PhysRevLett.118.054101,Rini2007,disa_photoinduced_2023,Basini2024}: pumping the IR-active phonon indirectly displaces the $A_{1g}$ dimerization mode through a symmetry-allowed cubic anharmonicity of the form $X_{\rm IR}^2 X_{\rm R}$. The interaction transferring energy between the two phonon branches is written as\begin{equation}
			H_{\rm IR-R} = \frac{\kappa}{2\sqrt{N}}\sum_{\bq, \bq'} \left({\qir}_{,\bq'} {\qir}_{,\bq} - Z^\IR_{\bq,\bq'}\right) {\qr}_{,-\bq-\bq'} \,,
		\end{equation}
		where $Z^\IR_{\bq,\bq'} = \expval{X_{\IR,\bq}X_{\IR,-\bq}}_{c,0} =\frac{\delta_{\bq+\bq'}}{2 m_\IR \omega_{\IR,\bq}}$ is the zero-point fluctuation of the IR-active phonon and acts as a counterterm ensuring that the ground state Raman amplitude expectation value is zero. The cubic interaction is parameterized by the dimensionless coupling constant $\kappa = \bar \kappa  (\sqrt{m_\R}\omega_\R m_\IR \omega_\IR^2)$.
		
		Below the spin-Peierls transition temperature, the chain spontaneously dimerizes into alternating strong ($J$) and weak ($J'$) exchange bonds~[Fig.~\ref{fig:schematic}, right], described by the Hamiltonian\begin{align} 	H_{\rm S} &= \sum_l \left(J \vec{S}_{1,l}\cdot \vec{S}_{2,l} + J' \vec{S}_{2,l}\cdot \vec{S}_{1,l+1}\right)\,. \end{align}To describe the magnetic degrees of freedom in this dimerized phase, we employ the bond-operator formalism~\cite{sachdev_bond-operator_1990, gopalan_spin_1994, kumar_bond_2010,PhysRevB.107.174415, PhysRevB.103.045132, c3tb-h5hv}. This framework parametrizes the local Hilbert space of each dimer via hard-core bosonic creation operators for the singlet ($s^\dagger$) and triplet ($t^\dagger_\alpha$, with $\alpha \in \{x,y,z\}$) states. These operators are subject to the strict local single-occupancy constraint $s^\dagger_l s_l + \sum_\alpha t^\dagger_{l,\alpha} t_{l,\alpha} = 1$ at every dimer site $l$. In terms of bond operators, the local spin operators take the form
		\begin{subequations}
			\begin{align}
				\vec S_{1,l} &= \frac{1}{2}\bigg(s_l^\dagger t_{l,\alpha} + s_l t^\dagger_{l,\alpha} -i \sum_{\beta, \eta} \varepsilon_{\alpha\beta\gamma} t^\dagger_{l,\beta} t_{l,\eta} \bigg) \vec e_\alpha \,,\\
				\vec S_{2,l} &= \frac{1}{2}\bigg(- s_l^\dagger t_{l,\alpha} - s_l t^\dagger_{l,\alpha} -i\sum_{\beta, \eta} \varepsilon_{\alpha\beta\eta} t^\dagger_{l,\beta} t_{l,\eta}\bigg) \vec e_\alpha \,,
			\end{align}
		\end{subequations}
		where $\vec e_\alpha$ denote the Cartesian unit vectors. Standard mean-field treatments enforce the single-occupancy constraint via a global Lagrange multiplier $\mu$ and a finite singlet condensate $\bar s = \expval{s}$~\cite{sachdev_bond-operator_1990}. Because self-consistent dynamical updates of these parameters are computationally prohibitive for nonequilibrium simulations, we adopt the strong-dimerization limit ($J \gg J'$), fixing $\bar s = 1$ and $\mu = -3J/4$. Neglecting the resulting quartic triplet interactions~\cite{kumar_bond_2010, PhysRevB.103.045132, c3tb-h5hv}, this quadratic approximation accurately describes the system dynamics for inter-dimer couplings up to $\lambda = J'/J \approx 0.5$~\cite{4ddn-y88c}, but inevitably breaks down near the isotropic limit $J' \to J$. Applying a spatial Fourier transform followed by a Bogoliubov rotation parameterized by $\tanh(2\theta_\bk) = \lambda \cos\bk / (2 - \lambda \cos\bk)$, the spin Hamiltonian is diagonalized into non-interacting triplon quasiparticles~\cite{PhysRevB.107.174415,PhysRevB.103.045132,c3tb-h5hv, PhysRevB.110.064420},
		\begin{align}
			H_{\rm S} &= \sum_\bk \omega_\bk, \tilde t_{\bk,\alpha}^\dagger \tilde t_{\bk,\alpha}\,.
		\end{align}
		Here, $\tilde t^\dagger_{\bk,\alpha}$ creates a triplon with the dispersion relation $\omega_\bk = J\sqrt{1 - \lambda \cos\bk}$, defining an excitation gap $\Delta = J\sqrt{1 - \lambda}$ above the singlet ground state.

		Furthermore, the Raman-active mode dynamically modulates the intra- and inter-dimer superexchange couplings along the chain via the spin-phonon interaction,
		\begin{align}H_{\rm R-S} = \sum_\bl {\qr}_{,\bl} \left(g\,\vec{S}_{1,l}\cdot \vec{S}_{2,l} + g'\,\vec{S}_{2,l}\cdot \vec{S}_{1,l+1}\right)\,.\end{align}We set $g' = - g$, reflecting that the dimerization mode oppositely affects the intra- and inter-dimer bonds (increasing $J$ while proportionately decreasing $J'$). We establish the stability criteria for the system under this specific coupling regime in Appendix~\ref{apa}. Transformed into the triplon basis, the zone-center spin-phonon interaction simplifies to
		\begin{align}
			H_{\rm R-S} = \frac{X_\R }{\sqrt{N}}\sum_\bk \left(\Ak n_\bk + \frac{\Bk}{2}\left(z_\bk + z_\bk^\dagger\right)\right)\,,
		\end{align}where $n_\bk = t_{\bk,\alpha}^\dagger t_{\bk,\alpha}$ is the triplon density and $z_\bk = t_{\bk,\alpha}^\dagger t_{-\bk,\alpha}^\dagger$ creates an $S=0$ bitriplon~\cite{4ddn-y88c}. The coupling vertices are defined as\begin{subequations} \begin{align} 	\Ak &= \frac{2gJ - (g'J + gJ')\cos\bk}{2\omega_\bk}\,, \\ 	\Bk &= \frac{(gJ' - g'J)\cos\bk}{2\omega_\bk}\,. \end{align} \end{subequations}This zone-center formulation represents the $\bq=0$ limit of the general momentum-space interaction $H_{\rm R-S} = \frac{1}{\sqrt{N}}\sum_{\bq,\bk} {\qr}_{,\bq} \left( A_{\bk,\bq} t^\dagger_{\bk,\alpha} t_{\bk - \bq,\alpha} + \frac{B_{\bk,\bq}}{2}t^\dagger_{\bk,\alpha} t^\dagger_{ \bq - \bk,\alpha} + \text{H.c.}\right)\,.$ 
		
		Combining these distinct sectors, the total Hamiltonian governing the driven system is $H =  H_{\rm IR} + H_{\rm l-IR} + H_{\rm R} + H_{\rm IR-R} + H_{\rm S} + H_{\rm R-S}$.
		
		\subsection{Equations of motion}
		We derive the equations of motion (EoM) for the driven dynamics using a second-order cumulant expansion (see Appendix~\ref{apb}). The cumulant hierarchy is closed by assuming that higher-order connected correlations are negligible relative to lower-order factorizations \cite{kubo_generalized_1962}, e.g.,
		\begin{align}
			\frac{1}{\sqrt{N}}\sum_\bq \expval{\qrq t^\dagger_{\bk,\alpha} t^\dagger_{\bk - \bq,\alpha}}_c \ll \expval{\bqr}\expval{z_\bk}\,.
		\end{align}The relaxation pathways arising from these neglected higher-order terms, alongside interactions with external environmental modes (such as acoustic phonons), are phenomenologically captured by the effective damping rates $\gamma_\R$, $\gamma_\IR$, and $\gamma_\s$. Dissipation is explicitly incorporated via the Heisenberg-picture Lindblad master equation, given by
		\begin{align}
			\ddt O(t) = i [H, O] + \sum_i \gamma_i \left( L_i^\dagger O L_i - \frac{1}{2}\{L_i^\dagger L_i, O\}\right)\,,
		\end{align}where, for simplicity, we assume a zero-temperature environment. This restricts the Lindblad jump operators $L_i$ strictly to the annihilation operators: $b_\R$, $b_\IR$, and $t_{\bk,\alpha}$. 
		
		Accordingly, the time evolution for the triplon density and anomalous bitriplon correlator is given by{\small\begin{subequations}\label{eq:triplon_EoM}
				\begin{align}
					\ddt \langle n_\bk\rangle = {} &-i B_\bk \expval{z_\bk - z^\dagger_\bk}\expval{\bqr} - \gamma_\s \expval{n_\bk}\,, \\ 	\ddt \langle z_\bk\rangle = {} &2i\left( \omega_\bk + A_\bk\expval{\bqr} \right)\expval{z_\bk} + i B_\bk(2\expval{n_\bk} + 3)\expval{\bqr} \notag \\ {} &- \gamma_\s \expval{z_\bk}\,. 
				\end{align} 
		\end{subequations}}For the macroscopic zone-center phonons, we introduce the intensive variables $\bar X = X/\sqrt{N}$ and $\bar P = P/\sqrt{N}$. Their mean-field trajectories are governed by\begin{subequations}\label{eq:phonon_MF_EoMs}
			\begin{align} 	
				\ddt \langle\bqir\rangle = {} &\frac{1}{m_\IR}\expval{\bpir} - \frac{\gamma_\IR}{2}\expval{\bqir}\,, \\ 
				\ddt \langle\bpir\rangle = {} &-m_\IR\omega_\IR^2 \expval{\bqir} - \frac{\gamma_\IR}{2}\expval{\bpir} + E(t) \notag \\ {} &- \kappa \left( \expval{\bqir} \expval{\bqr} + \expval{\qir\qr}_c\right)\,, \\ 	
				\ddt \langle\bqr\rangle = {} &\frac{1}{m_\R}\expval{\bpr} - \frac{\gamma_\R}{2}\expval{\bqr} \,,\\ 
				\ddt \langle\bpr\rangle ={} & -m_\R\omega_\R^2 \expval{\bqr} - \expval{\mathcal{E}} - \frac{\gamma_\R}{2}\expval{\bpr}\notag \\ {} & - \frac{\kappa}{2} \left( \expval{\bqir}^2 + \expval{\qir\qir}_c - Z^\IR\right), 
			\end{align} 
		\end{subequations}where we have dropped the momentum index for the uniform modes, and defined the effective spin-induced field on the lattice as $\mathcal{E} = \frac{1}{N}\sum_\bk\left(A_\bk n_\bk + \frac{B_\bk}{2}(z_\bk + z^\dagger_\bk)\right)$. 
		
		Due to the strong SPC, the Raman-active phonon experiences a significant renormalization of its eigenfrequency. From linear response theory, the Raman self-energy induced by the SPC is given by
		\begin{equation}
			\Pi_\R(\omega) = \frac{2}{N}\sum_k \frac{\tilde \omega_k B_k^2(2\expval{n_k} + 3)}{\omega^2 - (4\tilde \omega_k^2 + \gamma_{\rm s}^2) + 2i\gamma_{\rm s} \omega}\,,
		\end{equation}
		where $\tilde \omega_k = \omega_k + A_k \expval{\bar X_\R}$
		is the mean-field dynamical triplon dispersion. In equilibrium, the expression simplfies because $\expval{\bar X_\R} = 0$, leading to $\tilde\omega_k = \omega_k$ and $\expval{n_k} = 0$. The ground-state eigenfrequency of the dressed phonon is then given by the pole of its full propagator,
		\begin{equation}
			D_\R(\omega) = \frac{1}{\omega^2 - \omega_{R,\mathrm{bare}}^2-\Pi_\R(\omega) + i\gamma_\R \omega}\,.
		\end{equation}
		In the weak-damping, the renormalized (physical) resonance frequency $\omega_{\R,\mathrm{ren}}$ satisfies
		\begin{equation}
			\omega_{\R,\mathrm{ren}}^2 - \omega_{\R,\mathrm{bare}}^2 - \Re \Pi_\R(\omega_{\R,\mathrm{ren}}^2) = 0\,.
		\end{equation}
		Because only the renormalized resonance frequency $\omega_{\R,\mathrm{ren}}$ is experimentally accessible, we must ensure the model reproduces this physical value. To achieve this, we introduce a counter-term into the EoMs \eqref{eq:phonon_MF_EoMs} by setting the bare frequency to $\omega_\R = \sqrt{\omega_{\R,\mathrm{ren}}^2 - \Re \Pi_\R(\omega^2_{\R,\mathrm{ren}})}$,	which guarantees that the mode's dressed eigenfrequency in the ground state exactly matches $\omega_{\R,\mathrm{ren}}$. In principle, an additional self-energy correction arises from the phonon-phonon interaction; however, we assume the anharmonic coupling $\kappa$ is small compared to the SPC $g$, rendering this contribution negligible.
		
		Lastly, to capture the quantum fluctuations and nonlinear feedback, the connected two-point phonon correlators evolve according to a closed set of generalized Lyapunov equations. In the environment, these take the form\begin{widetext}
			\begin{subequations}
				\begin{align} 
					&	\ddt \langle \qir^2\rangle_c = {} \frac{1}{m_\IR}\langle \{\qir, \pir\}\rangle_c - \gamma_\IR \left(\expval{\qir^2}_c-\frac{1}{2m_\IR \omega_\IR}\right)\,, \\ 	
					&	\ddt \frac{1}{2}\langle \{\qir, \pir\}\rangle_c ={}  \frac{1}{m_\IR}\expval{\pir^2}_c - m_\IR \omega_\IR^2 \expval{\qir^2}_c - \kappa \left(\expval{\bqr}\expval{\qir^2}_c + \expval{\bqir}\expval{\qir \qr}_c\right)- \frac{\gamma_\IR}{2} \langle \{\qir, \pir \}\rangle_c\,, \\ 
					&	\ddt \langle \pir^2\rangle_c = {} - m_\IR \omega_\IR^2 \langle \{\qir, \pir \} \rangle_c - \kappa \left( \expval{\bqr} \langle \{\qir , \pir \} \rangle_c + 2 \expval{\bqir} \expval{\qr\pir}_c \right) - \gamma_\IR \left(\expval{\pir^2}_c-\frac{m_\IR \omega_\IR}{2}\right) \,, \\ 	
					&\ddt \langle \qr^2\rangle_c = {} \frac{1}{m_\R}\expval{\{\qr , \pr \} }_c - \gamma_\R\left(\expval{\qr^2}_c-\frac{1}{2m_\R\omega_\R}\right) \,, \\ 	
					&\ddt \frac{1}{2}\langle \{\qr,\pr\}\rangle_c = \frac{1}{m_\R}\expval{\pr^2}_c - m_\R\omega_\R^2\expval{\qr^2}_c-\kappa \expval{\bqir}\expval{\qr\qir}_c - \frac{\gamma_\R}{2}\langle \{\qr,\pr\} \rangle_c \,, \\ 	
					&\ddt \langle \pr^2\rangle_c = -m_\R\omega_\R^2\expval{\{\qr,\pr\}}_c - 2\kappa \expval{\bqir} \expval{\pr\qir}_c- \gamma_\R\left(\expval{\pr^2}_c -\frac{m_\R \omega_\R}{2}\right) \,, \\ 	
					&\ddt \langle \qr\qir \rangle_c = \frac{1}{m_\R}\expval{\pr\qir}_c + \frac{1}{m_\IR}\expval{\pir\qr}_c - \frac{\gamma_\IR + \gamma_\R}{2}\expval{\qir\qr}_c \,, \\ 	
					&\ddt \langle \pir \qr \rangle_c = \frac{1}{m_\R}\expval{\pr\pir}_c - m_\IR\omega_\IR^2\expval{\qir\qr}_c - \kappa \left( \expval{\bqr }\expval{\qr \qir}_c + \expval{\bqir }\expval{\qr^2}_c\right) - \frac{\gamma_\IR + \gamma_\R}{2}\expval{\pir\qr}_c\,,  \\ 
					&\ddt \langle \pr \qir \rangle_c = \frac{1}{m_\IR}\expval{\pr\pir}_c - m_\R\omega_\R^2\expval{\qir\qr}_c - \kappa \expval{\bqir}\expval{\qir^2}_c - \frac{\gamma_\IR + \gamma_\R}{2}\expval{\qir\pr}_c \,, \\ 
					&\ddt \langle \pir \pr \rangle_c = - m_\R\omega_\R^2 \expval{\qr \pir}_c - m_\IR\omega_\IR^2 \expval{\qir \pr}_c - \frac{\kappa}{2}\expval{\bqir}\left(\expval{\{\qr,\pr\}}_c + \expval{\{\pir,\qir\}}_c\right)  - \kappa \expval{\bqr}\expval{\qir\pr}_c \notag \\ 	&\qquad\qquad\qquad- \frac{\gamma_\IR + \gamma_\R}{2}\expval{\pir\pr}_c \,.
				\end{align} 
			\end{subequations}
		\end{widetext}

		\subsection{Energy flows}
		
		To rigorously quantify the instantaneous energy flows between interacting subsystems, we consider a generic bipartite system governed by the Hamiltonian $\hat H = \hat H_a + \hat H_b + \hat H_{ab}$. By evaluating the Heisenberg-picture Lindblad equations for the individual subsystem Hamiltonians, we can isolate the coherent energy transfer. Specifically, the power flowing from the interaction term into subsystem $b$ is given by the commutator
		\begin{equation}
			\mathcal{P}_{ab\to b}(t) = \frac{d}{dt}\expval{\hat H_b}(t) = i \expval{\left[\hat H_{ab},\hat H_b\right]}(t)\,.
		\end{equation}It should be emphasized that, even in the steady state, this coherent power transfer is not generally identical to the flow originating from subsystem $a$, $\mathcal{P}_{a\to ab}(t)$. This inequality arises because the total energy derivative of the subsystem also accounts for energy continuously dissipating into the environment via the non-unitary Lindblad channels.
        \begin{figure}
			\centering
			\includegraphics[width=0.9\linewidth]{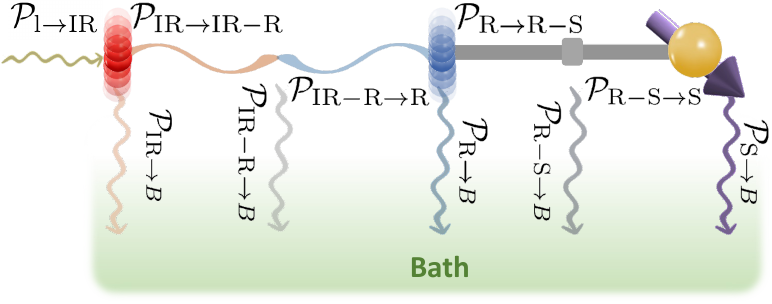}
			\caption{Schematic of cascaded energy flows and dissipative channels. Energy injected into the driven IR-active mode ($\mathcal{P}_{\mathrm{l}\to\mathrm{IR}}$) cascades sequentially through the cubic anharmonic interaction potential ($\mathcal{P}_{\mathrm{IR}\to\mathrm{IR-R}}$ and $\mathcal{P}_{\mathrm{IR-R}\to\mathrm{R}}$) into the Raman-active dimerization mode. The coherent Raman lattice displacement subsequently transfers power into the spin-phonon interaction potential ($\mathcal{P}_{\mathrm{R}\to\mathrm{R-S}}$) and into the spin subsystem ($\mathcal{P}_{\mathrm{R-S}\to\mathrm{S}}$) to generate triplon excitations. At each stage of the cascade, energy is continuously dissipated into the thermal environment via non-unitary Lindblad channels ($\mathcal{P}_{a\to\mathrm{B}}$ for subsystem $a \in$ \{IR, IR-R, R, R-S, S\}).}
			\label{fig:schematic2}
		\end{figure}

		Following the schematic in Fig.~\ref{fig:schematic2}, we first enumerate the energy flows within the system's closed sector. The flows associated with the open dissipative sector are collectively denoted as $a \rightarrow \text{B}$, indicating direct energy transfer to the bath.
		
		The rate of energy absorption by the IR-active phonon from the driving laser field is given by
		\begin{align}
			\mathcal{P}_{{\rm l}\to \IR}(t) = i\expval{[H_\mathrm{l-\IR},H_\IR]}(t) = \sqrt{N}E(t)\frac{\expval{P_\IR}(t)}{m_\IR}\,.
		\end{align}This defines the total input power injected into the system. Establishing this quantity provides a strict baseline against which we can evaluate the subsequent non-equilibrium energy distribution, allowing us to track the fractional routing of power through the anharmonic coupling and ultimately into the Raman and spin degrees of freedom.
		
		The coherent power transferred from the IR mode to the cubic interaction potential is evaluated as
		\begin{align}\label{eq_21}
			\mathcal{P}_{\IR \to {\rm IR-R}}(t) = {} & i\expval{[H_\IR,H_{\rm IR-R} ]}(t) \notag \\ = {} & \frac{\kappa}{2m_\IR}\expval{X_\R \left\{X_\IR,P_\IR \right\}}(t).
		\end{align}
		This demonstrates that the instantaneous energy flow is dynamically gated by the Raman amplitude $X_\R$ and the IR kinetic response $\{X_\IR, P_\IR\}$. Similarly, the coherent power transferred from the cubic interaction potential into the Raman phonon branch is given by
		\begin{align}
			\mathcal{P}_{{\rm IR-R} \to \R}(t) = {} &i\expval{[H_{\rm IR-R} ,H_\R]}(t) \notag \\ = {} & -\frac{\kappa}{2m_\R} \expval{X_\IR^2 P_\R}(t)\,.
		\end{align}Accordingly, the coherent power transferred from the Raman phonon into the spin-phonon interaction $H_{\rm R-S}$ is determined by
		\begin{equation}
			\mathcal{P}_{\R \to {\rm R-S}}(t) = i\expval{[H_\R,H_{\rm R-S}]}(t) = \frac{1}{m_\R}\expval{P_\R \mathcal{E}}(t)\,,
		\end{equation}
		where $\mathcal{E}$ represents the bare spin exchange operator coupling to the Raman lattice displacement.
		
		Finally, the coherent power transferred from the spin-phonon interaction potential into the bare spin subsystem is extracted via the commutator $i\expval{[H_{\rm R-S}, H_{\rm S}]}$. Because the interaction commutes with the particle-conserving terms of the triplon Hamiltonian, the energy transfer is driven entirely by the anomalous bitriplon creation and annihilation processes. This yields
		\begin{align}\label{eq_24}
			\mathcal{P}_{{\rm R-S} \to {\rm S}}(t) = {} & i\expval{[H_{\rm R-S},H_{\rm S}]}(t) \notag \\ = {} & \frac{2}{N} \sum_k \omega_k B_k \expval{X_\R \Im z_k}(t)\,.
		\end{align}This expression highlights that the continuous injection of energy into the magnetic excitations relies on the generation of an out-of-phase, imaginary component in the triplon correlator.
		
		The energy dissipated from the individual subsystems and interaction potentials into the thermal bath is determined by the negative of their respective damping terms in the Heisenberg-picture Lindblad equation, defined as
		\begin{equation}
			\mathcal{P}_{a\to B} = -\sum_i \gamma_i \left\langle\left( L_i^\dagger H_a L_i - \frac{1}{2}\{L_i^\dagger L_i, H_a\}\right)\right\rangle\,.
		\end{equation}Evaluating this for each component of the Hamiltonian yields the dissipation rates as{\small\begin{subequations}
			\begin{align}
				&\mathcal{P}_{\IR \to B}(t) = {} \gamma_\IR \omega_\IR \expval{n_\IR}(t)\, ,\\
				&\mathcal{P}_{{\rm IR-R} \to B}(t) = {}\left(\gamma_\IR + \frac{\gamma_\R}{2}\right)\expval{H_{\rm IR-R} }(t) -  \frac{\gamma_\IR\kappa\expval{X_\R}(t)}{4m_\IR \omega_\IR}\,,\\
				& \mathcal{P}_{\R \to B}(t) ={} \gamma_\R \omega_\R \expval{n_\R}(t)\,,\\
				&\mathcal{P}_{{\rm R-S} \to B}(t) = {}\left(\gamma_{\rm s} + \frac{\gamma_\R}{2}\right) \expval{X_\R \mathcal{E}}(t)\,,\\
				&\mathcal{P}_{{\rm S}\to B}(t) = {} \frac{\gamma_{\rm s} }{N}\sum_k \omega_k \expval{n_k}(t)\,.
			\end{align}
		\end{subequations}}To evaluate the phonon occupations $\expval{n_\IR}(t)$ and $\expval{n_\R}(t)$, we express the number operators in terms of the canonical coordinates and momenta for $j \in \{\IR, \R\}$ as
		\begin{align}
			\expval{n_j}(t) = \frac{m_j \omega_j}{2} \expval{X_j^2}(t) + \frac{1}{2m_j\omega_j} \expval{P_j^2}(t) - \frac{1}{2}\,.
		\end{align}Within the cumulant expansion framework, these second moments decouple into mean-field and fluctuation components (e.g., $\expval{X_j^2} = \expval{X_j}^2 + \expval{\delta X_j^2}$), allowing the dynamic populations to be computed directly from the solutions to the equations of motion.
		
		Notably, unlike the energy flows originating from the bare subsystems, the dissipation rates associated with the interaction potentials ($H_{\rm IR-R} $ and $H_{\rm R-S}$) are not positive definite. This is a direct consequence of utilizing the noninteracting annihilation operators as jump operators in the Lindblad dissipator. Because the interaction energy itself can be negative, its decay toward the noninteracting steady state strictly requires a compensatory energy flow from the bath back into the system.
		
		\section{Results and discussion}\label{s3}
		
		In earlier theoretical treatments of a simplified single-phonon architecture~\cite{PhysRevB.103.045132}, the driven IR mode was assumed to couple directly to the magnetic exchange interactions (a formulation reviewed in Appendix~\ref{apc}). In that direct-coupling scenario, the SPC constants $g$ and $g'$ are generally weak and share the same sign. Consequently, the efficiency of energy transfer into the spin sector is governed predominantly by the unperturbed triplon DOS, with absorption peaking strictly at the van Hove singularities of the band edges. 
		
		Channeling the energy cascade through the Raman-active dimerization mode alters this paradigm. Because the Raman mode physically modulates the alternating lattice dimerization, a dynamic displacement that compresses one nearest-neighbor bond simultaneously expands the adjacent one. The associated SPCs therefore inherently adopt opposite signs ($g' \approx -g$). This antisymmetric bond modulation dramatically amplifies the effective scattering vertex, $B_k \sim gJ' - g'J$, thrusting the system firmly into a strong-coupling regime. In this limit, immense dynamical back-action from the spin network becomes the rate-limiting quantity, forcing the peak energy transfer away from the van Hove singularities~(unperturbed bitriplon band edges) of the DOS.
		
		To construct a complete picture of the dynamical energy cascade, we first simulate the real-time evolution of the system following the onset of the THz drive. Evaluating the power flow through each sector requires simultaneously tracking multiple coupled observables. For instance, the energy transfer from the IR mode into the intermediate cubic potential [Eq.~\eqref{eq_21}] depends on the canonical coordinates of both phonons, while the power flow into the magnetic sector [Eq.~\eqref{eq_24}] is dictated by the macroscopic Raman displacement and the triplon pair correlations. We therefore begin by analyzing the temporal buildup of these foundational observables. Having established these real-time dynamics, the remainder of our discussion focuses on mapping the time-averaged energy flows, where the optical pumping is ultimately balanced by environmental dissipation.
		
		Throughout our calculations, parameters are systematically chosen near their physical upper bounds within the stable regime to maximize non-equilibrium signatures without triggering non-physical instabilities. The coupled dynamics of the multi-tiered phononic network and spin continuum define a high-dimensional parameter space—spanning drive amplitude ($E_0$), laser frequency ($\Omega_{\rm d}$), phonon frequencies ($\omega_\IR$, $\omega_\R$), anharmonicity ($\kappa$), exchange anisotropy ($J'$), SPCs ($g, g'$), and dissipation rates ($\gamma_{\text{IR}}, \gamma_{\text{R}}, \gamma_{\rm s}$). An exhaustive sweep across this manifold is computationally intractable due to the high dimensionality of the parameter space. Instead, we employ a physically grounded mapping strategy centered on realistic upper limits for material parameters. By probing these physical boundary limits, we ensure that the predicted signatures are robust, experimentally realizable phenomena rather than artifacts of a fine-tuned parameter set. 
		
		In practice, the scale of $\omega_\mathrm{R}$ is material-specific; for example, in $\mathrm{CuGeO}_3$, three principal dimerization modes at $6.05$, $10.56$, and $21.83~\mathrm{THz}$ were identified~\cite{jz36-8kz9}, corresponding roughly to $\omega_\mathrm{R} = 2J$, $\omega_\mathrm{R} = 4J$, and $\omega_\mathrm{R} = 8J$ assuming $J \approx 11.5~\mathrm{meV}$. Given the significantly larger value of $J \approx 60~\mathrm{meV}$ in $\mathrm{TiOCl}$~\cite{PhysRevLett.95.097203}, lower relative eigenfrequencies of the dimerization mode are expected in that material. From there, an IR-active mode with strong coupling to the desired dimerization mode must be identified. The ratio between their harmonic eigenfrequencies, $\omega_\mathrm{IR}$ and $\omega_\mathrm{R}$, dictates which regime of $\Omega_\mathrm{d}/\omega_\mathrm{R}$ can be driven efficiently. In the following, we focus on the resonant regime, namely $\omega_\mathrm{R} \approx 2\omega_\mathrm{IR} \approx 2\Omega_\mathrm{d}$, where maximum energy transfer into the spin sector occurs. Another compelling parameter configuration arises when $\omega_\mathrm{R} \ll \omega_\mathrm{IR} \approx \Omega_\mathrm{d}$. In this limit, difference-frequency generation arising from the anharmonic phonon-phonon coupling dominates the drive of the Raman mode compared to sum-frequency generation at $2\Omega_\mathrm{d}$. Because a static lattice displacement performs negligible work on the spin system, very few triplons are directly excited during the drive. Instead, the rectified Raman mode modifies the triplon dispersion by tuning the dynamical exchange couplings, $\tilde{J}(t) = J + g\langle X_\mathrm{R}\rangle(t)$, quasi-statically for the duration of the pump. Thus, the drive can dynamically enlarge ($\kappa g < 0$) or compress ($\kappa g > 0$) the spin gap depending on the relative sign of $\kappa$ and $g$. Under ultra-short pulse excitation, this scenario provides an attractive platform for ultrafast spintronic control.
		
		\subsection{Continuous-wave drive}
		Throughout this section, we model the external THz drive in Eq.~\eqref{eq_2} as a monochromatic CW field, $E(t) = E_0 \sin(\Omega_{\rm d} t)$. Figure~\ref{f2} tracks the real-time evolution of both the structural and magnetic observables as the system is driven toward the NESS. By explicitly tuning the Raman frequency near the bitriplon band edge ($\omega_{\rm R}/J = 1.45$), we investigate how strong resonant hybridization between the lattice and the magnetic continuum dictates the transient and steady-state dynamics.
		
		Upon the activation of the optical drive at $t=0$, the IR phonon displacement $X_\IR(t)$ grows rapidly at the fundamental frequency $\Omega_{\rm{d}}$. Because the structural resonance condition is strictly enforced ($\omega_\R = 2\omega_\IR = 2\Omega_{\rm{d}}$), the cubic anharmonicity ($\bar \kappa = -0.1$) acts as an efficient nonlinear frequency doubler, parametrically pumping the Raman mode at $2\Omega_{\rm{d}}$. While the resulting macroscopic amplitude of the Raman mode is substantially smaller than that of the primary IR phonon [necessitating the scaling in Fig.~\ref{f2}(a)], this subordinate displacement functions as the critical geometric bridge to the magnetic sector. This energy transfer is maximized by combining the strong antisymmetric SPC ($-\bar g' = \bar g = 0.6$) with a Raman frequency tuned near the bitriplon band edge ($\omega_{\rm R}/J = 1.45$). 
		
		By $t \approx 20$ ps, continuous optical injection balances the inherent lattice dissipation ($\gamma_{\R,\IR}$), and the structural observables stabilize into NESS orbits. Crucially, in this late-time limit, $X_\R(t)$ acquires a finite DC offset. This dynamic shift of the structural equilibrium position arises from two distinct rectifying forces. The first is an anharmonic optical rectification driven by the time-averaged non-zero mean of the primary mode ($\bar{\kappa}\langle X_\IR^2 \rangle < 0$). The second is a dynamic magnetostriction, which serves as a definitive signature of massive magnetic back-action. The steady-state accumulation of non-equilibrium triplons near the resonant band edge exerts a continuous, directional stress back onto the lattice, acting as a light-induced analogue of a spin-Peierls distortion to minimize the total free energy of the strongly driven system.\begin{figure}[t]
			\centering
			\includegraphics[width=0.9\linewidth]{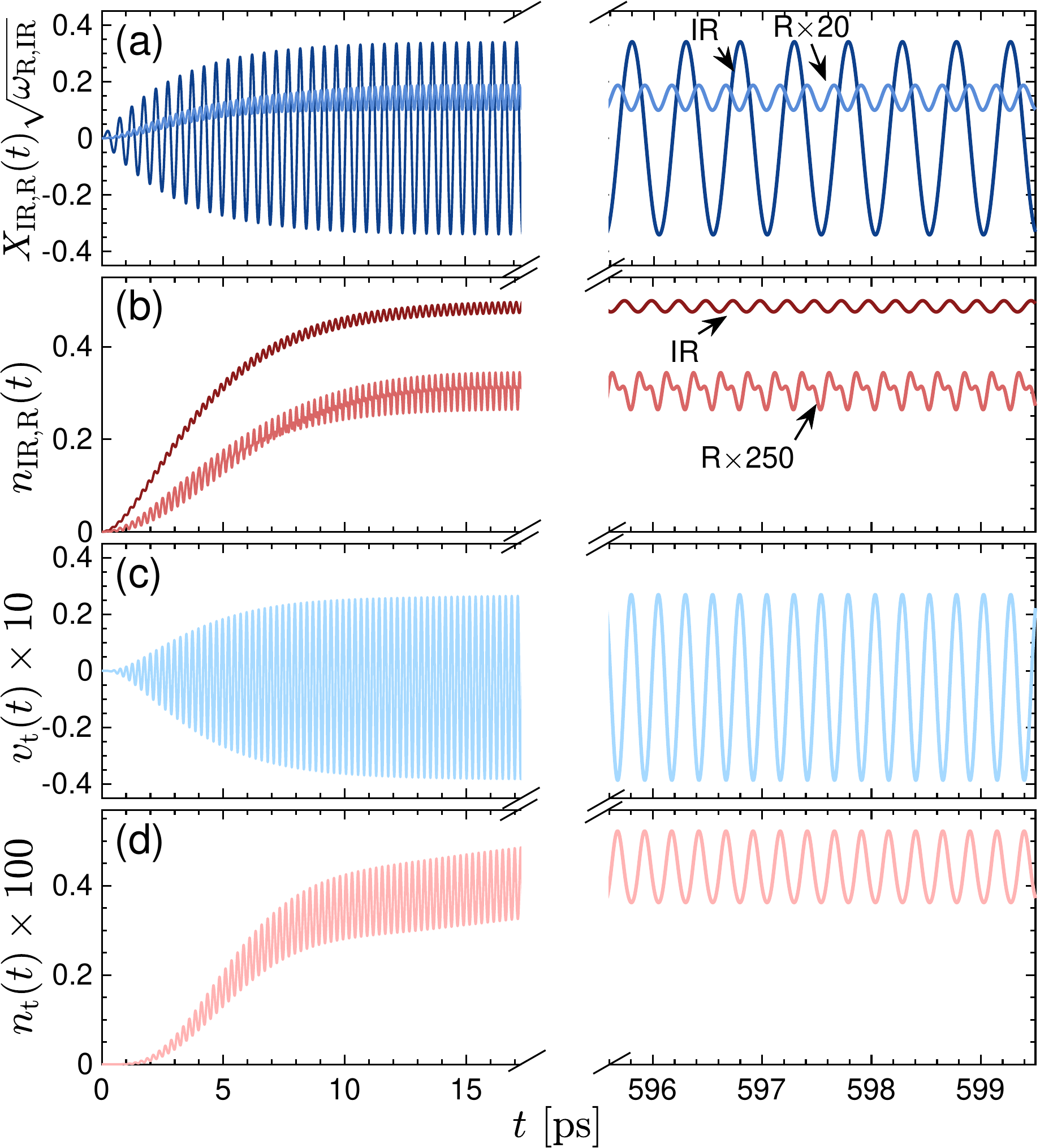}
			\caption{Temporal evolution of the driven observables from the initial turn-on of the CW THz field ($t = 0$, left panels) to the NESS (right panels). From top to bottom: (a) normalized canonical phonon displacements $X_{\IR,\R}(t)\sqrt{\omega_{\R,\IR}}$, (b) phonon populations $n_{\IR,\R}(t)$, (c) average triplon pair correlations $v_t(t) = \sum_k \expval{\Re( z_k)(t)} \slash N$, and (d) the average triplon population $n_t(t) = \sum_k \langle n_k(t)\rangle /N$. The system is driven at resonance $\omega_\R = 2\omega_\IR = 2\Omega_{\rm{d}}$, optimizing the nonlinear cubic anharmonicity $\bar \kappa = -0.1$. To maximize coupling to the magnetic sector, the Raman frequency is tuned near the bitriplon band edge, $\omega_{\rm R}/J = 1.45$. The antisymmetric SPCs are set to $-\bar g' = \bar g = 0.6$ with inter-dimer exchange $J' = J/2$. Bath dissipation rates are $\gamma_{\R,\IR} = 0.05\,\omega_{\R,\IR}$ and $\gamma_{\rm s} = 0.01\,J$, with a pump amplitude $E_0 = 0.05\,\omega_\IR^{3/2}$. For visibility, some of the observables are rescaled, as indicated.}
			\label{f2}
		\end{figure}
		
		While the canonical displacements track the coherent structural oscillations, the phonon populations $n_\IR(t)$ and $n_\R(t)$ [Fig.~\ref{f2}(b)] measure the time-averaged energy stored within each mode. Upon activation of the CW THz pump, the IR population $n_\IR(t)$ exhibits a rapid, massive accumulation. Because this mode couples directly to the external field, its population dominates the structural energy landscape, plateauing at a macroscopic steady state where optical injection balances bath dissipation ($\gamma_\IR$). In contrast, the Raman population $n_\R(t)$ relies on the nonlinear cubic anharmonicity ($\bar \kappa$) to siphon energy from the IR mode. Thus, its transient buildup shows a noticeable lag, and its steady-state magnitude remains substantially smaller. This behavior underscores the Raman mode's functional role: rather than acting as a primary energy reservoir, it serves as a high-frequency kinetic bridge to the spin sector, a pathway optimized by its proximity to the bitriplon band edge.
		
		Figures~\ref{f2}(c) and~\ref{f2}(d) capture the consequential excitation and back-action within the magnetic sector. Continuous modulation of the exchange interactions via $X_\R(t)$ drives the dynamic creation of magnetic excitations. The triplon pair correlation $v_t(t) = \sum_k \expval{\Re( z_k)(t)} \slash N$ oscillates symmetrically around zero, directly measuring the coherent creation and annihilation of triplon pairs induced by the oscillating dimerization field. The macroscopic accumulation of these excitations is tracked by the total triplon population $n_t(t) = \sum_k \langle n_k(t)\rangle/N$. Intimately tied to the magnetostrictive shift of the Raman coordinate, $n_t(t)$ acquires a pronounced DC offset that builds throughout the transient regime before plateauing at a limit set by the magnetic relaxation rate ($\gamma_{\rm s} = 0.01\,J$). In the late-time NESS limit, the population does not become strictly static; rather, it maintains $2\Omega_{\rm d}$ oscillations superimposed on the non-zero time-average. These ripples are a definitive signature of strong hybridization at the band-edge resonance, reflecting the continuous injection and extraction of energy within a single optical cycle.
		
		A striking feature of the population dynamics is the vast disparity in absolute magnitudes; $n_t(t)$ is scaled by a factor of 100 simply for visual comparison with $n_\IR(t)$. That the asymptotic triplon and Raman phonon populations remain orders of magnitude smaller than the IR-mode's population is a profound signature of the antisymmetric bond modulation ($g' \approx -g$). Because this specific symmetry and precise frequency matching amplify the effective scattering vertex to such an extreme degree, the triplon sector acts as an overwhelming load on the dimerization mode, almost completely stalling it. The work that the IR-mode can perform against the overdamped Raman mode is in turn diminished, choking the energy cascade already at the transition between the two phonons (see also Fig.~\ref{f3}).
		\begin{figure}[t]
			\centering
			\includegraphics[width=0.9\linewidth]{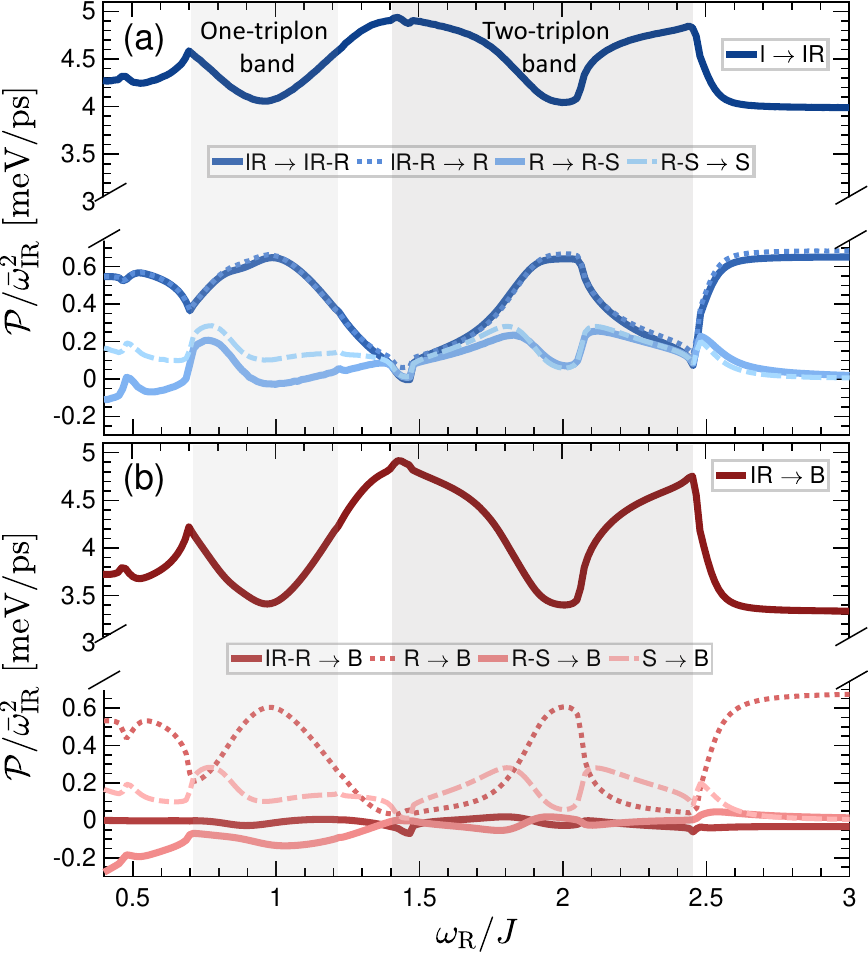}
			\caption{Spectral profile of the nonequilibrium energy flows in the driven spin-Peierls chain as a function of the Raman tuning $\omega_{\text{R}}/J$. (a, top) Total input power injected by the CW THz laser into the primary IR-active mode. (a, bottom) Coherent power transfer through the cascaded subsystems (IR $\to$ anharmonic coupling $\to$ Raman $\to$ SPC $\to$ bare spins). (b) Corresponding dissipation rates into the thermal bath, highlighting the non-positive-definite nature of the decay associated with the intermediate interaction potentials. The external drive is a CW field $E(t) = E_0 \sin(\Omega_{\rm d} t)$ with resonance strictly maintained at $\omega_\R = 2\omega_\IR = 2\Omega_{\rm{d}}$. Shaded regions indicate the extent of the one- and bitriplon excitation continua. Parameters are identical to those in Fig.~\ref{f2}.}
			\label{f3}
		\end{figure}
		
		Turning to a systematic analysis of the NESS, we find that the energy flow through the driven spin-Peierls chain exhibits a highly nontrivial dependence on the Raman tuning, $\omega_{\mathrm{R}}/J$. As illustrated in Fig.~\ref{f3}(a, top), the total optical power injected into the IR phonon, $\mathcal{P}_{{\rm l}\to{\rm IR}}$, is far from constant. Instead, the absorption profile displays distinct resonant peaks and pronounced suppressions that correlate directly with the energetic boundaries of the one- and bitriplon continua. This behavior is a hallmark of strong dynamical back-action. In highly resonant regimes, the IR mode can draw more power from the CW drive because the triplon sector stalls the Raman mode, effectively freezing one of the channels that create drag on the IR mode.
		
		The internal routing of this injected power is tracked via the commutator relations across the intermediate subsystems, as detailed in Fig.~\ref{f3}(a, bottom). The coherent power flowing from the IR mode into the cubic anharmonic potential ($\mathcal{P}_{\rm{IR}\to{\rm IR-R}}$) almost perfectly matches the subsequent transfer rate into the bare Raman mode ($\mathcal{P}_{{\rm IR-R}\to\rm{R}}$). This overlap indicates that the intermediate nonlinear lattice coupling introduces negligible dissipation to the energy cascade. However, severe attenuation occurs at the final interface between the lattice and the magnetic network. The power flowing from the bare Raman mode into the magnetoelastic coupling potential ($\mathcal{P}_{{\rm R}\to {\rm R-S}}$) drops significantly compared to the energy circulating within the pure phonon sector. Crucially, this bottleneck restricts the cascade not only at the band center—where the effective coupling $B_k$ trivially vanishes—but also at the band edges, where the coupling is maximized and the bitriplon DOS peaks. This confirms that peak energy transfer into the magnetic sector is not dictated strictly by the bare DOS. Rather, maximum throughput is governed by a stringent dynamical impedance-matching condition (detailed below), which requires optimal phase alignment between the coherent Raman distortion and the out-of-phase generation of the anomalous triplon density. Furthermore, at low frequencies, we observe a negative net energy flow from the Raman mode to the spin-phonon interaction potential, indicating a local reversal of the flow direction. In this regime, the interaction potential actively extracts energy from the thermal environment to stabilize the highly driven state, thereby reversing the gradient.
		
		Finally, Fig.~\ref{f3}(b) maps the continuous thermal dissipation into the environment via the Lindblad damping channels, a necessary mechanism for stabilizing the NESS against runaway heating. The IR mode overwhelmingly dominates the net heat dissipation ($\mathcal{P}_{\mathrm{IR}\to {\rm B}}$), effectively acting as a primary thermal short-circuit that shunts the majority of the injected laser power directly into the bath before it can propagate through the anharmonic coupling. 
		
		Beneath this dominant background, the subordinate dissipation channels reveal the microscopic footprint of the cascade's attenuation. Direct thermalization of the bare Raman mode ($\mathcal{P}_{\mathrm{R}\to {\rm B}}$) and the bare spin sector ($\mathcal{P}_{{\rm S}\to {\rm B}}$) proceeds at rates an order of magnitude lower than the IR dissipation, directly mirroring the severe energetic bottlenecks established in the coherent routing. The spin dissipation $\mathcal{P}_{{\rm S}\to {\rm B}}$, which is strictly proportional to the steady-state triplon density, serves as a reliable proxy for the surviving magnetic excitations. Notably, its peaks synchronize perfectly with the optimal dynamical impedance-matching conditions rather than the unperturbed triplon band edges. 
		
		Furthermore, the dissipation rates associated with the intermediate interaction potentials---specifically the anharmonic coupling ($\mathcal{P}_{{\rm IR-R} \to {\rm B}}$) and the spin-phonon interaction ($\mathcal{P}_{{\rm R-S}\to {\rm B}}$)---exhibit localized regimes of negative energy flow. The spin-phonon channel, in particular, displays pronounced negative valleys at lower Raman frequencies ($\omega_{\mathrm{R}}/J \lesssim 1.3$). Because the interaction energies are not intrinsically positive definite, their continuous decay toward the noninteracting steady state demands a compensatory influx of heat from the thermal environment to sustain the strongly driven state.
		
		To firmly establish the microscopic mechanism underlying the suppression of energy transfer at the band edges, the following provides a rigorous analytical treatment of the dynamical impedance-matching condition. Although impedance matching governs power transfer in classical electrodynamics, it is often overlooked in the context of driven quantum materials. This omission typically occurs because external optical fields are often treated as rigid, unyielding sources, or because the internal couplings are too weak to induce appreciable back-action between the interacting subsystems. To formalize this, we evaluate the system in the steady state. The instantaneous power deposited into the spin network by the Raman mode is given by
		\begin{equation}
			\mathcal{P}_{{\rm R-S} \to {\rm S}} = i\expval{[H_{\rm R-S},H_{\rm S}]} = \frac{2}{N} \sum_k \omega_k B_k \expval{X_\R \Im z_k}\,.
		\end{equation}
		
		We define the mean-field dynamical spin dispersion as
		\begin{equation}
			\tilde \omega_k = \omega_k + A_k \expval{X_\R}\,,
		\end{equation}
		and treat the triplon density $\expval{n_k}$ as approximately constant. As dictated by the density EoM, $\expval{n_k}$ primarily contains a rectified DC shift and a second-harmonic component relative to the phonon drive; thus, it performs no net work against the lattice displacement $X_\R$. The real and imaginary components of the anomalous correlator $z_k$ can then be obtained by treating the spin-phonon modulation $B_k (2\expval{n_k} + 3) X_\R(t)$ as an inhomogeneous driving field. This yields the linear response relations
		\begin{subequations}
			\begin{align}
				\expval{\Re z_k}(\omega) &= \chi_{1,k}(\omega) \expval{X_\R}(\omega)\,,\\
				\expval{\Im z_k}(\omega) &= \chi_{2,k}(\omega) \expval{X_\R}(\omega)\,,
			\end{align}
		\end{subequations}
		which are governed by the respective susceptibilities
		\begin{subequations}\begin{align}
				\chi_{1,k}(\omega) &= \frac{2\tilde \omega_k B_k(2\expval{n_k}+3)}{\omega^2 - \left(4\tilde \omega_k^2 + \gamma_{\rm s}^2\right) + 2i\omega\gamma_{\rm s}}\,, \\
				\chi_{2,k}(\omega) &= \frac{\left(i \omega - \gamma_{\rm s} \right) B_k(2\expval{n_k}+3)}{\omega^2 - \left(4\tilde \omega_k^2 + \gamma_{\rm s}^2\right) + 2i\omega\gamma_{\rm s}}\,.
			\end{align}
		\end{subequations}
		
		Similarly, the EoM for the Raman phonon can be recast in frequency space as a single second-order equation,
		\begin{equation}
			D_0^{-1}(\omega) \expval{\bar{X}_\R}(\omega) = E(\omega) + \expval{\mathcal{E}}(\omega)\,,
		\end{equation}
		where the inverse bare phonon propagator is given by
		\begin{equation}
			D_0^{-1}(\omega) = \omega^2 - \tilde \omega_\R^2 + i \omega \gamma_\R\,,
		\end{equation}
		with $\tilde \omega_\R^2 = \omega_\R^2 + \gamma_\R^2/4$. Here, $E(\omega)$ captures the effective driving force mediated by the IR-active mode. The dynamic spin back-action, $\mathcal{E}$, can be partitioned into a phonon self-energy,
		\begin{equation}
			\frac{1}{N}\sum_k B_k \expval{\Re z_k}(\omega) = \Pi_\R(\omega) \expval{\bar{X}_\R}(\omega)\,,
		\end{equation}
		where
		\begin{equation}
			\Pi_\R(\omega) = \frac{1}{N}\sum_k B_k \chi_{1,k}(\omega)\,,
		\end{equation}
		and a static rectification source, $\frac{1}{N}\sum_k A_k \expval{n_k}$, which we absorb into the effective drive $E(\omega)$ as it does not contribute to the dynamic impedance. Notice that this is the magnetostrictive force-inducing part of the DC-shift displayed by $\expval{X_\R}$ in Fig.~\ref{f2}.
		
		The fully dressed Raman phonon response is then solved via
		\begin{align}
			\expval{\bar{X}_\R}(\omega) = {} &\frac{E(\omega)}{\omega^2 - \left(\tilde \omega_\R^2 +  \Pi_\R'(\omega)\right) + i \left( \omega \gamma_\R - \Pi_\R''(\omega)\right)}\notag \\ \equiv {} & \chi_\R(\omega) E(\omega)\,.
		\end{align}
		The dissipative, imaginary part of the self-energy evaluates to
		\begin{equation}
			\Pi_\R''(\omega) = - \frac{4\omega \gamma_{\rm s}}{N}\sum_k \frac{ \tilde \omega_k B_k^2 (2\expval{n_k}+3)}{\left[\omega^2 - \left(4\tilde \omega_k^2 + \gamma_{\rm s}^2\right)\right]^2 + 4\omega^2\gamma_{\rm s}^2}\,.
		\end{equation}
		Notably, in the limit of vanishing spin damping ($\gamma_{\rm s} \to 0$), this term reduces to the bitriplon DOS weighted by the square of the SPC $B_k$:
		{\small\begin{align}
				\Pi_\R''(\omega) \xrightarrow{\gamma_{\rm s} \to 0} - \mathrm{sgn}(\omega)\frac{2\pi}{N}\sum_k  B_k^2 (2\expval{n_k}+3) \tilde\omega_k \delta\left(\omega^2 - 4\tilde \omega_k^2 \right).
		\end{align}}For finite $\gamma_{\rm s}$, the effect is qualitatively identical, with the only modification being that the bitriplon DOS appears smeared out $\sim \gamma_{\rm s}$ to the dimerization mode.
		If we consider a harmonic drive of the form $E(t) = E_0 \cos(\Omega_{\rm d} t)$, the average steady-state power transfer (the zero-frequency component) per unit cell becomes
		\begin{equation}
			\bar{\mathcal{P}}_{{\rm R-S}\to {\rm S}} = E_0^2 \abs{\chi_\R(\Omega_{\rm d})}^2 \frac{1}{N}\sum_k B_k \omega_k \, \chi_{2,k}'(\Omega_{\rm d})\,.
		\end{equation}
		Evaluating this in the $\gamma_{\rm s} \to 0$ limit simplifies the extraction of the resonance physics. The imaginary susceptibility yields{\small\begin{equation}
				\chi'_{2,k}(\Omega_{\rm d}) \to   2\tilde \omega_k \pi\delta(\Omega_{\rm d}^2 - 4\tilde \omega_k^2) B_k (2\expval{n_k}+3)\,,
		\end{equation}}which, when inserted into the average power expression, produces
		\begin{equation}
			\bar{\mathcal{P}}_{{\rm R-S}\to {\rm S}} \approx  \frac{E_0^2 \Omega_{\rm d}^2}{2} \abs{\chi_\R(\Omega_{\rm d})}^2  \Gamma_{\rm s}(\Omega_{\rm d})\,,
		\end{equation}
		where we have defined the effective spin-induced damping rate{\small\begin{align}
				\Gamma_{\rm s}(\Omega_{\rm d}) = {} & \frac{\pi}{N}\sum_k \delta(\Omega_{\rm d}^2 - 4\tilde \omega_k^2) B_k^2 (2\expval{n_k}+3) = -\frac{\Pi_\R''(\Omega_{\rm d})}{\Omega_{\rm d}}
		\end{align}}and approximated $2 \omega_k \tilde\omega_k \delta(\Omega_{\rm d}^2 - 4\tilde\omega_k) \approx \Omega_d^2\slash 2$. Expanding the full response function, the total steady-state power flow evaluates to
		{\small\begin{equation}
				\bar{\mathcal{P}}_{{\rm R-S}\to {\rm S}} = \frac{1}{2}\frac{ E_0^2 \Omega_{\rm d}^2 \Gamma_{\rm s}(\Omega_{\rm d})}{\left[\Omega_{\rm d}^2 - \left(\tilde \omega_\R^2 +  \Pi_\R'(\Omega_{\rm d})\right)\right]^2 + \Omega_{\rm d}^2 \left(  \gamma_\R + \Gamma_{\rm s}(\Omega_{\rm d})\right)^2}\,.
		\end{equation}}Maximizing this energy throughput with respect to the dynamic spin back-action $\gamma_{\rm s}(\Omega)$ yields the exact classical impedance-matching condition \cite{thompson2009dynamo},
		\begin{equation}
			\Gamma_{\rm s}^2(\Omega_{\rm d}) = \frac{\left[\Omega_{\rm d}^2 - \left(\tilde \omega_\R^2 +  \Pi_\R'(\Omega_{\rm d})\right)\right]^2}{\Omega_{\rm d}^2} + \gamma_\R^2\,.
		\end{equation}
		
		This rigorous analytical condition demonstrates that maximum energy throughput into the bare spins does not strictly coincide with the band edges. Instead, the cascade is optimized precisely when the dynamical drag exerted by the spin sector is perfectly matched to the intrinsic structural impedance of the lattice. If the back-action on the phonon is neglected---or equivalently, if the triplons are driven by a rigid, unyielding source---the power drawn by the spins scales proportionally with their effective coupling to the drive, $\Gamma_{\rm s}(\Omega_{\rm d})$. However, by properly retaining the dynamical back-action, we observe a critical turnover: if the effective spin drag $\Gamma_{\rm s}(\Omega_{\rm d})$ exceeds the intrinsic resistance of the phonon ($\gamma_\R$), the spins effectively stall the structural oscillation and thereby \emph{reduce} the net power transferred into the magnetic sector.
		
		For the parameter regime considered in Ref.~\cite{PhysRevB.103.045132}, the phonon-bitriplon coupling $B_k$ is sufficiently weak such that the phonon experiences negligible drag from the spins (i.e., $\Gamma_{\rm s}(\Omega_{\rm d}) \ll \gamma_\R$). In this weak-coupling limit, the absorbed power is naturally maximized by tuning the driving and phonon frequencies exactly to the band edge, where the bitriplon DOS, $B_k^2$, and consequently $\Gamma_{\rm s}(\Omega_{\rm d})$ are all maximized. In stark contrast, the physical dimerization mode with $g' = -g$ generates a massive $B_k$. Because $B_k$ crosses zero at the band center, the existence of an optimally impedance-matched $\Gamma_{\rm s}(\Omega_{\rm d})$ is guaranteed somewhere between the band center and the band edge. For larger maximal $B_k$, this optimum strictly shifts closer to the band center.\begin{figure}
			\centering
			\includegraphics[width=0.9\linewidth]{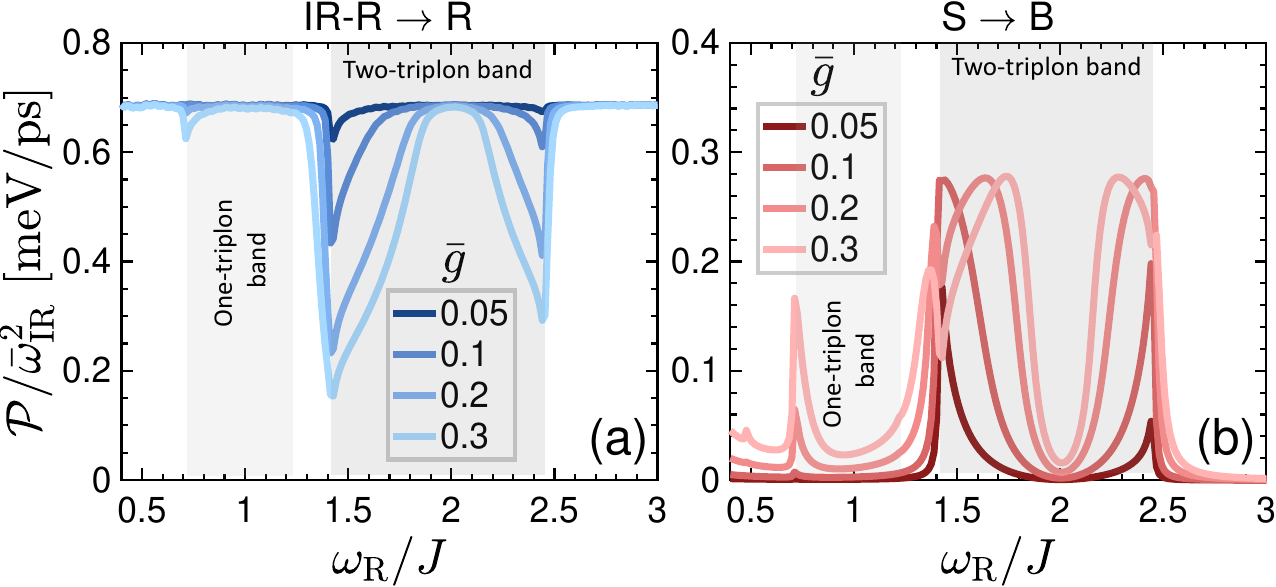}
			\caption{Steady-state average power transfer (a) from the anharmonic phonon interaction into the Raman mode ($\mathcal{P}_{{\rm IR-R}\to\rm R}$) and (b) from the spin sector into the thermal bath ($\mathcal{P}_{{\rm S}\to\rm B}$) versus Raman frequency tuning $\omega_\R/J$. Results are compared across the weak- to strong-coupling transition for three SPC strengths~($\bar g = 0.1,\, 0.2$, and $0.3$). In all cases, $\bar g' = -\bar g$, $\bar \kappa = -0.1$, $J' = J/2$, $\omega_\R = 2\omega_\IR = 2\Omega_{\rm{d}}$, $\gamma_{\IR/\R} = 0.05\,\omega_{\IR/\R}$, $\gamma_{\rm s} = 0.01\,J$, and $E = 0.05\,\omega_\IR^{3/2}$.}
			\label{f6}
		\end{figure}
		
		In addition to the strong SPC ($\bar g = -\bar g' = 0.6$) analyzed in Figs.~\ref{f2} and~\ref{f3}, we track a range of coupling strengths in Fig.~\ref{f6} to delineate the crossover between the weak- and strong-coupling regimes. Because most energy channels exhibit qualitatively similar dynamical impedance-matching behavior, we focus specifically on power transfer within the nonlinear phononic sector ($\mathcal{P}_{{\rm IR-R}\to\rm R}$) and dissipation from the magnetic sector into the thermal bath ($\mathcal{P}_{{\rm S}\to\rm B}$). Although the antisymmetric SPC structure ($g' \approx -g$) of the dimerization mode generally favors the strong-coupling regime by enhancing the effective phonon-bitriplon vertex $B_k \propto (g J' - g' J)$, weak-coupling behavior is naturally recovered for small values of $g$. As shown in Fig.~\ref{f6}, tuning the dimensionless SPC parameter $\bar g = -\bar g'$ from $0.05$ to $0.3$ drives a smooth crossover from the weak- to the strong-coupling regime. The primary signature of this transition manifests in the energy transfer rates from the Raman mode into the triplon sector—namely, the coherent injection $\mathcal{P}_{\mathrm{R}\to\mathrm{R\text{-}S}}$, the subsequent transfer $\mathcal{P}_{\mathrm{R\text{-}S}\to\mathrm{S}}$, and the resulting bath dissipation $\mathcal{P}_{\mathrm{S}\to\mathrm{B}}$ [Fig.~\ref{f6}(b)]. For $\abs{\bar g} \leq 0.1$, peak energy transmission occurs precisely at the bitriplon band edge; as the coupling increases to $\bar g = 0.3$, this peak shift provides a clear fingerprint of the crossover from DOS-dominated to impedance-matching-dominated energy transfer. Furthermore, the magnitude of maximum power transfer already saturates near the boundary of the weak-coupling regime ($\bar g = 0.1$). 
		
		Further increasing the coupling strength leaves the peak transmission amplitude invariant and only shifts its spectral position inside the bitriplon band. At this point, the bottleneck of the energy cascade into the spin sector moves up a rung from the spin-phonon to the phonon-phonon interaction. Within the phonon sector, the enhanced coupling between the triplon bath and the dimerization mode suppresses energy transfer from the primary IR mode into the Raman mode near the band edges, where the Raman mode becomes dynamically stalled [Fig.~\ref{f6}(a)]. Consequently, the primary IR mode experiences reduced back-action damping, enabling it to absorb more power from the laser drive and dissipate it directly into the thermal bath. In contrast to the energy transfer into the spin sector, the effect that increasing SPC has on the phonon sector only saturates at much larger couplings, once the energy transmission to the Raman mode at the band-edges reaches strictly zero [cf. Fig. \ref{f3}(a) and Fig. \ref{f6}(a)].
		
		\subsection{Pulsed drive}
		In this section, we parameterize the external THz drive introduced in Eq.~\eqref{eq_2} as a transient pulsed field, $E(t) = E_0 \sin(\Omega_{\rm d} t) s(t)$. The temporal profile is modulated by a Gaussian envelope, $s(t) = \exp\left[-(t - t_0)^2/2\tau^2\right]$, where $t_0$ marks the peak arrival time and $\tau$ characterizes the pulse duration. This duration is related to the full width at half maximum, $2\tau\sqrt{2\ln(2)}$. For all time-averaged observables evaluated in this transient regime, the temporal average is computed strictly over this effective pulse width; specifically, the results presented below reflect data time-averaged over a 10-period pulse.
		
		To disentangle the intrinsic resonant properties of the spin-Peierls lattice from the highly nonlinear saturation effects induced by prolonged optical pumping, we directly compare the time-averaged power spectra of the CW driven NESS with those of the transient pulsed regime [Figs.~\ref{f3} and \ref{f4}]. While both protocols exhibit the same fundamental Fano-like interference signatures near the triplon band edges, their spectral topographies differ drastically in absolute magnitude, modulation contrast, and linewidth. Fundamentally, this disparity arises from the interaction timescales. The CW field continuously injects energy, forcing the system to indefinitely maintain macroscopic populations of both phonons and triplons against the thermal bath. In contrast, averaging over the transient Gaussian envelope naturally incorporates the rising and falling tails of the pulse. This keeps the integrated energy density strictly perturbative, thereby circumventing the massive accumulation of excitations characteristic of the CW NESS limit.
		
		This disparity in excitation density fundamentally dictates the macroscopic back-action exerted on the primary lattice mode. In the CW regime, the modulation of the total absorbed optical power ($\mathcal{P}_{{\rm l}\to \IR}$) is colossal, varying by $20\%$ to $25\%$ as the Raman frequency is tuned across the bitriplon band edges [Fig.~\ref{f3}(a)]. Under pulsed excitation, this modulation contrast collapses to roughly $1\%$. Because the nonlinear back-action requires finite time to cascade through the anharmonic and magnetoelastic couplings, a transient pulse terminates before the spin sector can accumulate enough population to trigger a massive impedance shift. Consequently, the magnetic continuum merely perturbs the pulsed absorption spectrum, whereas it strictly dictates the CW response. Accordingly, we find strongly suppressed energy dissipation from the spins ($\mathcal{P}_{{\rm S}\to {\rm B}}$), both in absolute and relative terms, indicating extremely low triplon occupations in the pulsed drive scenario when compared to the CW drive.\begin{figure}[t]
			\centering
			\includegraphics[width=0.9\linewidth]{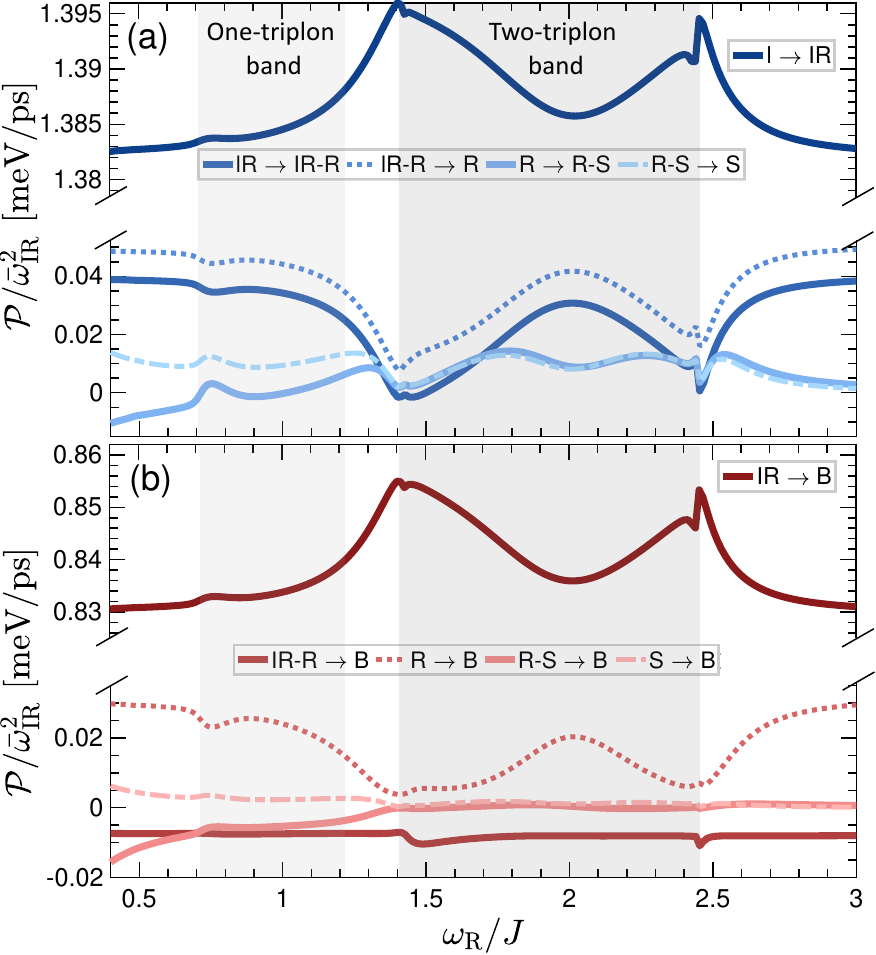}
			\caption{Same as Fig.~\ref{f3}, but replacing the CW drive with a transient Gaussian pulse, $E(t) = E_0 \sin(\Omega_{\rm d} t) s(t)$, having a full width at half maximum of ten pulse periods ($\tau_\mathrm{FWHM} = 10\,T_d$).}
			\label{f4}
		\end{figure}
		
		Comparing the spectral line shapes further reveals classic signatures of power broadening. The CW drive establishes a macroscopic steady-state triplon population that acts as a dense, hot background. The continuous, high-amplitude recycling of energy between the lattice and this excited spin bath introduces severe effective damping, smearing the abrupt band-edge responses into wide, continuous peaks and valleys. In contrast, the weak, transient nature of the pulsed drive circumvents this saturation, effectively probing the cold lattice and preserving the exquisitely sharp Fano-like interference features at the boundaries of the one- and bitriplon continua.
		
		Finally, the thermodynamic demands of the two driving protocols reveal a stark contrast in their dissipative routing. In the CW NESS limit, the spin-phonon injection ($\mathcal{P}_{\mathrm{R} \to \mathrm{R\text{-}S}}$) and its corresponding bath dissipation ($\mathcal{P}_{\mathrm{R\text{-}S} \to \mathrm{B}}$) exhibit deep regions of negative energy flow at low Raman frequencies ($\omega_\R/J \lesssim 1.3$). Maintaining the permanent macroscopic shifts in the lattice equilibrium positions (dynamic magnetostriction) far from equilibrium mandates this active heat extraction from the thermal environment. In the pulsed regime, these negative dissipative valleys are strongly suppressed, as the transient field simply does not drive the system deeply enough into the highly nonlinear regime to require such sustained thermodynamic extraction.
		
		\section{Summary and outlook}\label{s4}
		In this work, we have established a comprehensive theoretical framework to simulate non-equilibrium spin-phonon energy flows in low-dimensional quantum magnets. Taking the broken-symmetry phase of a strongly dimerized spin-Peierls chain as a concrete benchmark, we tracked the transient propagation and steady-state dissipation of power through a multi-tiered nonlinear phononics architecture. By combining a bond-operator representation of the $S=1/2$ magnetic sector with a second-order cumulant expansion of the Heisenberg-Lindblad equations of motion, our approach treats open-system relaxation and quantum fluctuations across thermodynamic limits without the severe spatial truncations inherent to exact diagonalization.
		
		The central physical insight emerging from our simulations is the breakdown of the unperturbed triplon DOS as the sole dictator of optical energy absorption. When light-driven energy cascades through an intermediate $A_{1g}$ Raman dimerization mode, the underlying SPC adopts an inherently antisymmetric vertex structure. This structural asymmetry amplifies the effective scattering channels, thrusting the system into a strong-coupling regime dominated by dynamical back-action. Consequently, energy throughput into the spin network is governed by a classical-like dynamical impedance-matching condition: maximum power transfer occurs not at the van Hove singularities of the bare triplon continua, but rather at frequencies where the dynamical drag exerted by the excited spin bath perfectly matches the intrinsic structural dissipation of the lattice.
		
		Crucially, the structural, spectral, and thermodynamic responses depend strongly on the drive's temporal profile. Continuous-wave illumination drives the system into a non-equilibrium steady state with a dense, saturated triplon population. This persistent spin background acts as a strong dissipative sink—inducing power broadening that blurs continuum edges and heavily dampens the primary infrared mode—while driving a static lattice distortion via dynamic magnetostriction. Thermodynamically, sustaining this macroscopic shift requires continuous entropy export, observed as negative power flows across system-bath interfaces. Conversely, a short pulse limits total fluence, keeping the response strictly perturbative. Because energy transfer across anharmonic and magnetoelastic channels requires finite time, the optical field decays before significant spin back-action can build or shift the lattice impedance. By probing an unheated spin network, the pulse avoids power broadening, preserving sharp Fano resonances and continuum interference dips without the heavy dissipative footprint of continuous driving.
		
		Beyond clarifying the non-equilibrium physics of a strongly dimerized spin-Peierls chain, our results offer important guidelines for the broader field of ultrafast magnetophononic control. First, identifying back-action-dominated regimes demonstrates that increasing laser intensity or coupling strength does not monotonically enhance spin excitation; exceeding the optimal impedance-matching threshold causes the spin network to dynamically choke the driving lattice mode, suppressing net power absorption. Accounting for this dynamical drag is essential for optimizing light-induced magnetic phase transitions and high-harmonic spin generation. Second, our thermodynamic accounting highlights how transient pulsed architectures can be leveraged to manipulate magnetic order parameter dynamics while mitigating the parasitic heating and active environmental heat demands that typically destabilize driven steady states. Future extensions of this framework could incorporate finite-temperature Lindblad channels to explore thermal phase boundaries, or generalize the cumulant expansion to encompass non-zero momentum ($q \neq 0$) acoustic phonon dynamics and multi-mode phonon interference.
		
		\section*{Acknowledgments}
		M.\,Y. and J.\,K.\,F. were supported by the Department of Energy, Office of Basic Energy Sciences, Division of Materials Sciences and Engineering under Contract No.\ DE-FG02-08ER46542 for the formal developments, the analytical/numerical work, and the writing of the manuscript. J.\,K.\,F. was also supported by the McDevitt bequest at Georgetown University.
		
		\section*{Data availability}
		The data that support the findings of this article are openly available at~\cite{Zenodo}.
		
		\appendix
		
		\section{Stability}\label{apa}
		To establish the stable parameter regime and the equilibrium configuration, it is instructive to apply a mean-field (MF) decoupling to the triplon Hamiltonian. Replacing the Raman phonon operator with its expectation value, the effective MF Hamiltonian is given by{\small\begin{align}
				H_{\rm S}^\text{MF} = \sum_\bk\Big(\left(\omega_\bk + A_\bk\expval{X_\R}\right) n_\bk + B_\bk\expval{X_\R} \Re z_\bk\Big)\,.
		\end{align}}This Hamiltonian is diagonalized via a Bogoliubov transformation parametrized by the mixing angles $\theta_\bk$, where
		\begin{align}
			\tanh 2\theta_\bk = - \frac{B_\bk\expval{X_\R}}{\omega_\bk + A_\bk\expval{X_\R}}\,.
		\end{align}This yields the renormalized triplon dispersion,
		\begin{align}
			\tilde \omega_\bk = \sqrt{(\omega_\bk + A_\bk\expval{X_\R})^2 - (B_\bk\expval{X_\R})^2}\,.
		\end{align}
		
		To ensure a real, positive excitation spectrum ($\tilde \omega_\bk > 0$), we require $\omega_\bk + A_\bk\expval{X_\R} > \abs{B_\bk\expval{X_\R}}$. For the specific case of $g' = -g$, this restricts the Raman phonon amplitude to
		\begin{equation}
			g\expval{X_\R} > - \frac{J-J'}{2}
		\end{equation}
		to maintain a stable triplon spectrum at the MF level. This constraint inherently reflects the validity limits of the dimerized singlet ansatz; violating it signals an instability toward the gapless uniform spin-chain regime, which is characterized by a continuum of two-spinon excitations. The ground-state expectation values for $n_\bk$ and $z_\bk$ are evaluated by requiring the new quasiparticle vacuum to be annihilated by $\gamma_{\bk,\alpha}$, where the original triplon operators are given by $t_{\bk,\alpha} = \cosh(\theta_\bk)\gamma_{\bk,\alpha} + \sinh(\theta_\bk) \gamma^\dagger_{\bk,\alpha}$. This yields
		\begin{subequations}
			\begin{align}
				\langle 0 \vert n_\bk\vert 0 \rangle = {} & \frac{3}{2}\left( \frac{\omega_\bk + A_\bk \expval{X_\R}}{\tilde \omega_\bk} - 1\right)\,, \\
				\langle 0 \vert z_\bk\vert 0 \rangle = {} & - \frac{3}{2}\frac{B_\bk\expval{X_\R}}{\tilde \omega_\bk}\,.
			\end{align}
		\end{subequations}
		
		Imposing the static equilibrium condition $\langle \dot{P}_\R \rangle = 0$ provides the balance equation
		\begin{equation}
			m_\R \omega_\R^2 \expval{X_\R} = - \frac{\kappa}{2}\left(\expval{X_\IR^2} - \frac{1}{2m_\IR \omega_\IR}\right) - \expval{\mathcal{E}}\,.
		\end{equation}Without external driving, the symmetry-breaking first term vanishes by construction.
		Hence, the self-consistency equation for the MF ground-state Raman amplitude becomes\begin{widetext}    
			\begin{align}
				m_\R \omega_\R^2 \expval{X_\R} = {} -\frac{3}{2N_k} \sum_\bk\left(\frac{A_\bk(\omega_\bk + A_\bk\expval{X_\R}) - B^2_\bk\expval{X_\R}}{\sqrt{(\omega_\bk + A_\bk\expval{X_\R})^2 - (B_\bk\expval{X_\R})^2}} - A_\bk\right)\,.
		\end{align}\end{widetext}The underlying physics of this renormalized state is clarified by defining effective exchange couplings, $\tilde J = J + g\expval{X_\R}$ and $\tilde J' = J' + g' \expval{X_\R}$. Expressed in terms of these shifted couplings, the dispersion takes the standard form
		\begin{equation}
			\tilde \omega_\bk = \tilde J \sqrt{1- \frac{\tilde J'}{\tilde J}\cos\bk}\,,
		\end{equation}
		and the self-consistency condition simplifies to
		\begin{equation}\label{eq:qR_self_consistency}
			0 = m_\R \omega_\R^2 \expval{X_\R} + \frac{3}{2N_k} \sum_\bk\left(\tilde A_\bk - A_\bk\right)  \,,
		\end{equation}
		where $\tilde A_\bk$ retains the functional form of $A_\bk$ but is evaluated using the renormalized couplings $\tilde J$ and $\tilde J'$. Because $\tilde A_\bk$ evaluates to $A_\bk$ given $\expval{X_\R} = 0$, the latter is guaranteed to be a solution of Eq.~\eqref{eq:qR_self_consistency}. To determine that this solution is stable, the right-hand side of Eq. \ref{eq:qR_self_consistency}, denoted by $f\left(\expval{X_\R}\right)$, must have a positive derivative at $\expval{X_\R} = 0$. We find
		\begin{equation}\label{eq:fprime}
			f'\left(\expval{X_\R}\right) = m_\R\omega_\R^2 - \frac{3g^2(J + J')}{8 N_k}\sum_\bk \frac{\cos^2(k)}{\tilde \omega_k^3}\,,
		\end{equation}
		where we exploited the fact that $\tilde J + \tilde J' = J + J'$. Importantly, the second term is strictly positive and only depends on $\expval{X_\R}$ via $\tilde \omega_k$. Thus the zero-crossing of Eq.~\eqref{eq:fprime} as a function of the SPC $g$ establishes a further stability criterion of the theory.
		
        \section{Cumulant expansion}\label{apb}
		Exact numerical simulation of large quantum systems is generally intractable due to the exponential growth of the Hilbert space. To address this, we use the cumulant expansion method (CEM)~\cite{kubo_generalized_1962,van_kampen_cumulant_1974}. The $n$th-order cumulant isolates genuine $n$-body correlations by systematically subtracting all permutations of lower-order factorizations from the full correlator. For example, the two- and three-point cumulants are defined as
		\begin{subequations}\label{eq:cumulants}
			\begin{align} 	\expval{\Oa\Ob}_c ={}& \expval{\Oa\Ob} - \expval{\Oa}\expval{\Ob}\,, \\ 	
				\expval{\Oa\Ob\Oc}_c ={}& \expval{\Oa\Ob\Oc} - \expval{\Oa}\expval{\Ob \Oc}_c - \expval{\Ob}\expval{\Oa \Oc}_c \notag \\ {} & - \expval{\Oc}\expval{\Oa \Ob}_c - \expval{\Oa}\expval{\Ob}\expval{\Oc}\,.
			\end{align} 
		\end{subequations}Truncating this expansion at order $n$ neglects $(n+1)$-body correlations, yielding a closed, computationally tractable set of EoM. While rigorous convergence bounds for the CEM are often unavailable~\cite{kerber2025cumulantsexpansionapproachgood}, its validity heavily depends on expanding around a suitable quasiparticle basis where higher-order correlations naturally remain small. Given our parameter regime of weak nonlinear phonon coupling ($\abs{\bar \kappa} \lesssim 0.1$) and moderate SPC ($\bar g \approx 0.5$), the bare, non-interacting phonon and triplon bases provide a justified starting point. For significantly stronger interactions ($\bar g \gtrsim 1$), expanding around hybridized phonon-bitriplon modes would become necessary~\cite{PhysRevB.107.174415}.
		
		To simplify the triplon equations, we evaluate the thermodynamic scaling of the phonon-triplon cross-correlations. Because the dimerized phase features a gapped excitation spectrum, the system possesses a finite correlation length $\xi$. Thus, local real-space correlations decay exponentially as{\small\begin{align}
			\expval{{\qr}_{,\bl} \, t^\dagger_{\bl',\alpha} t_{\bl'',\alpha}}_c \sim e^{-\frac{\abs{\bl' - \bl}}{\xi}-\frac{\abs{\bl'' - \bl}}{\xi}}\,.
		\end{align}}Summing over $\bl'$ and $\bl''$ for a fixed site $\bl$ converges to an intensive, $\mathcal{O}(1)$ value. Fourier transforming into momentum space extracts a $1/\sqrt{N}$ prefactor, yielding
		\begin{widetext}\begin{align}
				\expval{{\qr}_{,\bq} t^\dagger_{\bk,\alpha} t_{\bk - \bq,\alpha}}_c = \frac{1}{\sqrt{N}} \sum_{\bl',\bl''} e^{-i (\bl'-\bl) \bk + i (\bl'' - \bl)(\bk-\bq)} \expval{{\qr}_{,\bl} \, t^\dagger_{\bl',\alpha} t_{\bl'',\alpha}}_c \sim \mathcal{O}\left(\frac{1}{\sqrt{N}}\right)\,.
		\end{align}\end{widetext}This $\mathcal{O}(1/\sqrt{N})$ scaling holds universally for all connected correlators between the phonon coordinates ($\qr, \pr$) and triplon observables ($n, z$).
		
		The EoM for the triplon observables can be substantially simplified. We illustrate this using the density operator $n_{\bp,\bq} = \sum_\alpha t^\dagger_{\bp,\alpha} t_{\bq,\alpha}$, though identical considerations applied to the bitriplon operators $z^{(\dagger)}_{\bp,\bq}$. The full operator EoM reads
		\begin{widetext}\begin{subequations}\label{eq:general_momentum_n_EoM}
				\begin{align}
					\ddt n_{\bp,\bq} &= i(\omega_\bp - \omega_\bq) n_{\bp,\bq} - \gamma_{\rm s} n_{\bp,\bq}+ \frac{i}{\sqrt{N}}\sum_{\bk}X_{\R,\bk}\Big( A_{\bk+\bp,\bk} n_{\bk+\bp,\bq} - A_{\bq,\bk} n_{\bp,\bq-\bk}  - (B_{\bq+\bk,\bk} + B_{\bq,\bk}) z_{\bp,\bk-\bq}\notag \\
					&\quad  + (C_{-\bp-\bk,\bk} + C_{\bp,\bk}) z^\dagger_{\bq,-\bk-\bp} \Big)\,.
				\end{align}
		\end{subequations}\end{widetext}By enforcing translation invariance, finite expectation values are strictly bounded to diagonal observables, $n_\bk \equiv n_{\bk,\bk}$ and $z_\bk \equiv z_{\bk,-\bk}$. The exact EoM for the diagonal triplon density expectation value evaluates to\begin{widetext}\begin{align}
				\ddt \expval{n_{\bk}} = {}& - \gamma_{\rm s} \expval{n_{\bk}} + \frac{i}{\sqrt{N}}\sum_{\bq} \Big( A_{\bq+\bk,\bq} \expval{X_{\R,\bq}n_{\bq+\bk,\bk}} - A_{\bk,\bq} \expval{X_{\R,\bq}n_{\bk,\bk-\bq}} - (B_{\bk+\bq,\bq} + B_{\bk,\bq}) \expval{X_{\R,\bq}z_{\bk,\bq-\bk}} \notag \\ 	& + (C_{-\bk-\bq,\bq} + C_{\bk,\bq}) \expval{X_{\R,\bq}z^\dagger_{\bk,-\bq-\bk}} \Big)\,. \end{align}\end{widetext}For finite momentum transfers ($\bq \neq 0$), the mean-field factorizations trivially vanish. Consequently, these summands are driven entirely by the cubic connected cumulants (e.g., $\expval{X_{\R,\bq} n_{\bq + \bk,\bk}}_c$). 
			
			While the momentum sum over these individually $\mathcal{O}(1/\sqrt{N})$ cumulants formally scales as $\mathcal{O}(1)$, their distinct $\bq$-dependent oscillation frequencies—dictated by the finite triplon dispersion—drive rapid dephasing and destructive interference. Over timescales determined by the inverse triplon bandwidth, these fluctuations average to zero. Truncating the expansion by neglecting the $\bq \neq 0$ terms entirely collapses the sum, seamlessly recovering the simplified mean-field equation given in Eq.~\eqref{eq:triplon_EoM}. The same arguments apply identically for the anomalous correlators $z^{(\dagger)}_{\bp,\bq}$.
		
			The operator EoM for the phonons read\begin{subequations}\label{eq:canonical_phonon_EoMs}
				\begin{align}
					\frac{d}{dt}\pirq ={} & - \omega_\IR^2 \qirq  - \sqrt{N} E(t) \delta_{\bq,0} \notag \\ {}& - \frac{\kappa}{\sqrt{N}} \sum_{\bq'} X_{\IR,\bq + \bq'} X_{\R,-\bq'}\,,\\
					\frac{d}{dt}\prq = {} &- \omega_\R^2 \qrq  - \sqrt{N}\mathcal{E}_\bq \notag \\ {} & - \frac{\kappa}{2\sqrt{N}} \sum_{\bq'} \left(X_{\IR,\bq + \bq'} X_{\IR,-\bq'} - Z^\IR_{\bq, \bq'}\right) \,,
				\end{align}
		\end{subequations}where $\mathcal{E}_\bq$ is the generalized driving force originating from the full momentum-space SPC\begin{align}
				\mathcal{E}_\bq = {} &\frac{1}{N}\sum_\bk \Bigg( A_{\bk,\bq} t^\dagger_{\bk,\alpha} t_{\bk - \bq,\alpha}  + \frac{B_{\bk,\bq}}{2}t^\dagger_{\bk,\alpha} t^\dagger_{ \bq - \bk,\alpha}\notag \\ {} &+ \frac{B_{\bk, -\bq}^*}{2} t_{\bk,\alpha} t_{-\bq - \bk,\alpha}\Bigg)\,.
			\end{align}
		
		Because the laser drive is extensive, the zone-center expectation values scale macroscopically ($\sim \sqrt{N}$). Dividing Eq.~\eqref{eq:canonical_phonon_EoMs} by $\sqrt{N}$ yields the EoMs for the intensive (per-unit-cell) phonon operators, denoted by overbars{\small\begin{subequations}\label{eq:averaged_phonon_EoMs}
				\begin{align}
					\frac{d}{dt}\bar P_{\IR,\bq} = {} &- \omega_\IR^2 \bar X_{\IR,\bq} - \frac{\kappa}{N} \sum_{\bq'}  X_{\IR,\bq + \bq'}  X_{\R,-\bq'} - E(t) \delta_{\bq,0}\,,\\
					\frac{d}{dt}\bar P_{\R,\bq} = {} &- \omega_\R^2 \bar X_{\R,\bq} - \frac{\kappa}{2N} \sum_{\bq'} \left( X_{\IR,\bq + \bq'} X_{\IR,-\bq'} - Z^\IR_{\bq,\bq'}\right) \notag \\ {} & - \mathcal{E}_\bq\,.
				\end{align}
		\end{subequations}}Taking the expectation value of the interaction term ($\propto \kappa$) separates it into mean-field and connected contributions. To illustrate this, let us focus on the phonon-phonon interaction term in the IR mode EoM;
		\begin{equation}
			\left(\frac{1}{N}\sum_{\bq'}\expval{X_{\IR,\bq'} X_{\R,-\bq'}}_c + \expval{\bar X_{\IR}} \expval{\bar X_{\R}} \right)\delta_{\bq,0}\,.
		\end{equation}where, owing to translation invariance, only the zone-center mode expectation values are affected by the interaction. The connected term is not trivially negligible. The macroscopic zone-center mode is renormalized by the quantum fluctuations of $N$ phonon modes via momentum-conserving scattering. In Eq.~\eqref{eq:averaged_phonon_EoMs}, the sum over $N$ intensive fluctuations, each of $\mathcal{O}(1)$, is tempered by the $1/N$ prefactor from the definition of $\bar X$, placing it on equal footing with the mean-field amplitudes. Neglecting the connected term requires the coherent amplitude per unit cell to vastly exceed the covariance, $\expval{X^2}_c \ll \expval{\bar X}^2$. This condition is satisfied only under sufficiently strong driving—establishing a coherent state with per-unit-cell occupation $\bar n \gg 1$ that dominates vacuum fluctuations—and weak anharmonic coupling, $\bar \kappa \ll 1$. The first requirement, $\bar n \gg 1$, is not satisfied in our case.\begin{figure}[t]
			\centering
			\includegraphics[width=0.9\linewidth]{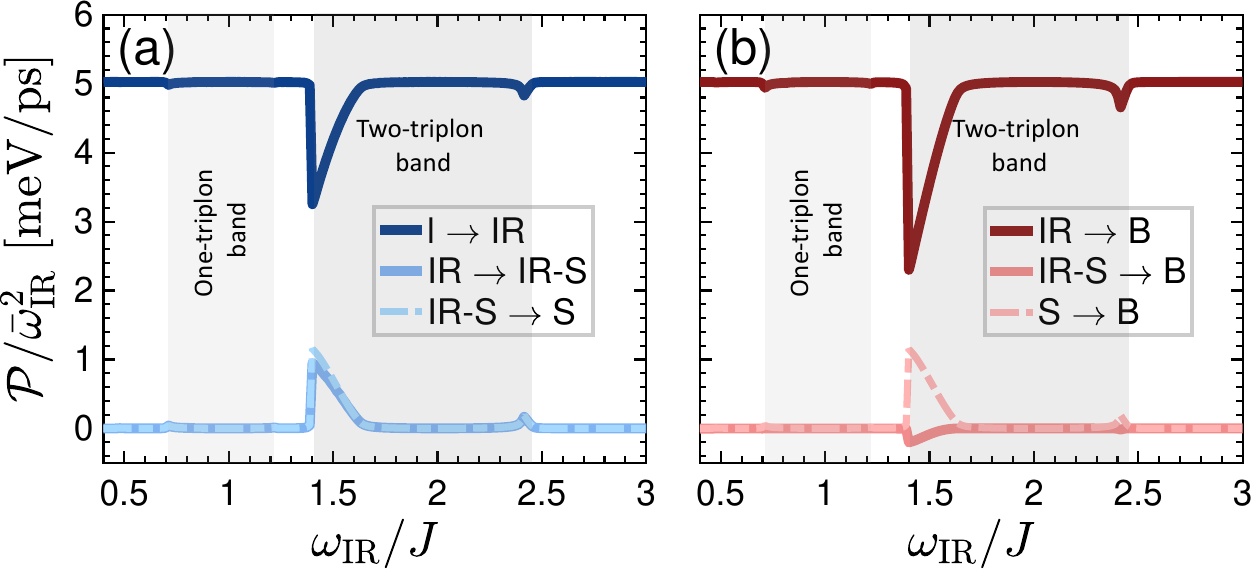}
			\caption{(a) Average power transfer versus driving frequency tuning $\omega_{\rm IR}/J$, from the laser to the IR phonon ($\mathcal{P}_{{\rm l} \to {\rm IR}}$), cascading through the intermediate interaction ($\mathcal{P}_{{\rm IR} \to {\rm IR-S}}$) into the magnetic sector ($\mathcal{P}_{{\rm IR-S} \to {\rm S}}$). Absorption peaks localize precisely at van Hove singularities at the one- and bitriplon band edges. (b) Thermal dissipation rates into the environment from the IR mode ($\mathcal{P}_{{\rm IR} \to {\rm B}}$), intermediate coupling ($\mathcal{P}_{{\rm IR-S} \to {\rm B}}$), and spin sector ($\mathcal{P}_{{\rm S} \to {\rm B}}$). Unlike the strong-coupling Raman case, energy routing in this direct-coupling scenario is strictly dominated by the bare density of states (DOS) with negligible dynamical back-action. Parameters: $\bar g' = 0.15$, $\bar g = 0.25$, $J' = J/2$, $\omega_\IR = \Omega_{\rm{d}}$, $\gamma_{\IR} = 0.05\,\omega_{\IR}$, $\gamma_{\rm s} = 0.01\,J$, and $E = 0.05\,\omega_\IR^{3/2}$.}
			\label{f5}
		\end{figure} 
		
		\section{Review of the single-phonon direct coupling model~\cite{PhysRevB.103.045132}}\label{apc}
		
		To contextualize the strong-coupling phenomena presented in the main text, we briefly review the energy flow within a simplified architecture where the driven IR phonon modulates the magnetic exchange directly. Because the structural displacements in this configuration lack the antisymmetric character of the Raman dimerization mode, the SPC parameters ($g$ and $g'$) typically share the same sign, yielding a significantly reduced effective scattering vertex. The steady-state energy currents for this direct-coupling scenario are presented in Fig.~\ref{f5}(a).
		
		Strikingly, the power injected into the system ($\mathcal{P}_{{\rm l} \to {\rm IR}}$) and its subsequent routing into the magnetic excitations ($\mathcal{P}_{{\rm IR-S} \to {\rm S}}$) exhibit sharp, well-defined resonances that align perfectly with the one- and bitriplon band edges. Because the effective spin-induced drag remains negligible compared to the intrinsic phonon damping ($\gamma_{\rm s} \ll \gamma_\IR$), the lattice dynamics remain effectively insulated from any magnetic back-action. In this weak-coupling limit, the triplon network acts as a passive reservoir, and the energy absorption profile is strictly governed by the bare DOS. This DOS-driven behavior is equally apparent in the system's thermalization pathways [Fig.~\ref{f5}(b)]. 
		
		The dissipation spectrum is heavily dominated by the direct relaxation of the IR phonon into the bath ($\mathcal{P}_{{\rm IR} \to {\rm B}}$), which acts as a massive thermal shunt. The subordinate dissipation channels from the interaction potential ($\mathcal{P}_{{\rm IR-S} \to {\rm B}}$) and the bare spins ($\mathcal{P}_{{\rm S} \to {\rm B}}$) closely mirror the absorption peaks, maximizing exactly at the van Hove singularities.
		
		Comparing these results to the multi-tiered Raman cascade underscores a shift in the underlying physics. When energy flows through a dimerization mode with antisymmetric couplings ($g' \approx -g$), the system transitions into a regime where the spin network exerts a massive dynamical drag on the lattice. It is this pronounced back-action that ultimately breaks the strict DOS reliance observed in Fig.~\ref{f5}, replacing it with the dynamical impedance-matching condition required to optimize energy transfer in strongly coupled quantum materials.
	}
	\bibliography{bib.bib}

\begin{thebibliography}{53}%
\makeatletter
\providecommand \@ifxundefined [1]{%
 \@ifx{#1\undefined}
}%
\providecommand \@ifnum [1]{%
 \ifnum #1\expandafter \@firstoftwo
 \else \expandafter \@secondoftwo
 \fi
}%
\providecommand \@ifx [1]{%
 \ifx #1\expandafter \@firstoftwo
 \else \expandafter \@secondoftwo
 \fi
}%
\providecommand \natexlab [1]{#1}%
\providecommand \enquote  [1]{``#1''}%
\providecommand \bibnamefont  [1]{#1}%
\providecommand \bibfnamefont [1]{#1}%
\providecommand \citenamefont [1]{#1}%
\providecommand \href@noop [0]{\@secondoftwo}%
\providecommand \href [0]{\begingroup \@sanitize@url \@href}%
\providecommand \@href[1]{\@@startlink{#1}\@@href}%
\providecommand \@@href[1]{\endgroup#1\@@endlink}%
\providecommand \@sanitize@url [0]{\catcode `\\12\catcode `\$12\catcode
  `\&12\catcode `\#12\catcode `\^12\catcode `\_12\catcode `\%12\relax}%
\providecommand \@@startlink[1]{}%
\providecommand \@@endlink[0]{}%
\providecommand \url  [0]{\begingroup\@sanitize@url \@url }%
\providecommand \@url [1]{\endgroup\@href {#1}{\urlprefix }}%
\providecommand \urlprefix  [0]{URL }%
\providecommand \Eprint [0]{\href }%
\providecommand \doibase [0]{https://doi.org/}%
\providecommand \selectlanguage [0]{\@gobble}%
\providecommand \bibinfo  [0]{\@secondoftwo}%
\providecommand \bibfield  [0]{\@secondoftwo}%
\providecommand \translation [1]{[#1]}%
\providecommand \BibitemOpen [0]{}%
\providecommand \bibitemStop [0]{}%
\providecommand \bibitemNoStop [0]{.\EOS\space}%
\providecommand \EOS [0]{\spacefactor3000\relax}%
\providecommand \BibitemShut  [1]{\csname bibitem#1\endcsname}%
\let\auto@bib@innerbib\@empty
\bibitem [{\citenamefont {Buzzi}\ \emph {et~al.}(2018)\citenamefont {Buzzi},
  \citenamefont {Först}, \citenamefont {Mankowsky},\ and\ \citenamefont
  {Cavalleri}}]{Buzzi2018}%
  \BibitemOpen
  \bibfield  {author} {\bibinfo {author} {\bibfnamefont {M.}~\bibnamefont
  {Buzzi}}, \bibinfo {author} {\bibfnamefont {M.}~\bibnamefont {Först}},
  \bibinfo {author} {\bibfnamefont {R.}~\bibnamefont {Mankowsky}},\ and\
  \bibinfo {author} {\bibfnamefont {A.}~\bibnamefont {Cavalleri}},\ }\bibfield
  {title} {\bibinfo {title} {Probing dynamics in quantum materials with
  femtosecond {X}-rays},\ }\href {https://doi.org/10.1038/s41578-018-0024-9}
  {\bibfield  {journal} {\bibinfo  {journal} {Nature Reviews Materials}\
  }\textbf {\bibinfo {volume} {3}},\ \bibinfo {pages} {299} (\bibinfo {year}
  {2018})}\BibitemShut {NoStop}%
\bibitem [{\citenamefont {de~la Torre}\ \emph {et~al.}(2021)\citenamefont
  {de~la Torre}, \citenamefont {Kennes}, \citenamefont {Claassen},
  \citenamefont {Gerber}, \citenamefont {McIver},\ and\ \citenamefont
  {Sentef}}]{delaTorre2021}%
  \BibitemOpen
  \bibfield  {author} {\bibinfo {author} {\bibfnamefont {A.}~\bibnamefont
  {de~la Torre}}, \bibinfo {author} {\bibfnamefont {D.~M.}\ \bibnamefont
  {Kennes}}, \bibinfo {author} {\bibfnamefont {M.}~\bibnamefont {Claassen}},
  \bibinfo {author} {\bibfnamefont {S.}~\bibnamefont {Gerber}}, \bibinfo
  {author} {\bibfnamefont {J.~W.}\ \bibnamefont {McIver}},\ and\ \bibinfo
  {author} {\bibfnamefont {M.~A.}\ \bibnamefont {Sentef}},\ }\bibfield  {title}
  {\bibinfo {title} {Colloquium: Nonthermal pathways to ultrafast control in
  quantum materials},\ }\href {https://doi.org/10.1103/RevModPhys.93.041002}
  {\bibfield  {journal} {\bibinfo  {journal} {Rev. Mod. Phys.}\ }\textbf
  {\bibinfo {volume} {93}},\ \bibinfo {pages} {041002} (\bibinfo {year}
  {2021})}\BibitemShut {NoStop}%
\bibitem [{\citenamefont {Afanasiev}\ \emph {et~al.}(2021)\citenamefont
  {Afanasiev}, \citenamefont {Hortensius}, \citenamefont {Ivanov},
  \citenamefont {Sasani}, \citenamefont {Bousquet}, \citenamefont {Blanter},
  \citenamefont {Mikhaylovskiy}, \citenamefont {Kimel},\ and\ \citenamefont
  {Caviglia}}]{Afanasievetal2021}%
  \BibitemOpen
  \bibfield  {author} {\bibinfo {author} {\bibfnamefont {D.}~\bibnamefont
  {Afanasiev}}, \bibinfo {author} {\bibfnamefont {J.~R.}\ \bibnamefont
  {Hortensius}}, \bibinfo {author} {\bibfnamefont {B.~A.}\ \bibnamefont
  {Ivanov}}, \bibinfo {author} {\bibfnamefont {A.}~\bibnamefont {Sasani}},
  \bibinfo {author} {\bibfnamefont {E.}~\bibnamefont {Bousquet}}, \bibinfo
  {author} {\bibfnamefont {Y.~M.}\ \bibnamefont {Blanter}}, \bibinfo {author}
  {\bibfnamefont {R.~V.}\ \bibnamefont {Mikhaylovskiy}}, \bibinfo {author}
  {\bibfnamefont {A.~V.}\ \bibnamefont {Kimel}},\ and\ \bibinfo {author}
  {\bibfnamefont {A.~D.}\ \bibnamefont {Caviglia}},\ }\bibfield  {title}
  {\bibinfo {title} {Ultrafast control of magnetic interactions via
  light-driven phonons},\ }\href
  {https://doi.org/https://doi.org/10.1038/s41563-021-00922-7} {\bibfield
  {journal} {\bibinfo  {journal} {Nature Materials}\ }\textbf {\bibinfo
  {volume} {20}},\ \bibinfo {pages} {607} (\bibinfo {year} {2021})}\BibitemShut
  {NoStop}%
\bibitem [{\citenamefont {Mitrano}\ \emph {et~al.}(2024)\citenamefont
  {Mitrano}, \citenamefont {Johnston}, \citenamefont {Kim},\ and\ \citenamefont
  {Dean}}]{Mitrano2024}%
  \BibitemOpen
  \bibfield  {author} {\bibinfo {author} {\bibfnamefont {M.}~\bibnamefont
  {Mitrano}}, \bibinfo {author} {\bibfnamefont {S.}~\bibnamefont {Johnston}},
  \bibinfo {author} {\bibfnamefont {Y.-J.}\ \bibnamefont {Kim}},\ and\ \bibinfo
  {author} {\bibfnamefont {M.~P.~M.}\ \bibnamefont {Dean}},\ }\bibfield
  {title} {\bibinfo {title} {Exploring quantum materials with resonant
  inelastic {X}-ray scattering},\ }\href
  {https://doi.org/10.1103/PhysRevX.14.040501} {\bibfield  {journal} {\bibinfo
  {journal} {Phys. Rev. X}\ }\textbf {\bibinfo {volume} {14}},\ \bibinfo
  {pages} {040501} (\bibinfo {year} {2024})}\BibitemShut {NoStop}%
\bibitem [{\citenamefont {Roelcke}\ \emph {et~al.}(2024)\citenamefont
  {Roelcke}, \citenamefont {Kastner}, \citenamefont {Graml}, \citenamefont
  {Biereder}, \citenamefont {Wilhelm}, \citenamefont {Repp}, \citenamefont
  {Huber},\ and\ \citenamefont {Gerasimenko}}]{Roelcke2024}%
  \BibitemOpen
  \bibfield  {author} {\bibinfo {author} {\bibfnamefont {C.}~\bibnamefont
  {Roelcke}}, \bibinfo {author} {\bibfnamefont {L.~Z.}\ \bibnamefont
  {Kastner}}, \bibinfo {author} {\bibfnamefont {M.}~\bibnamefont {Graml}},
  \bibinfo {author} {\bibfnamefont {A.}~\bibnamefont {Biereder}}, \bibinfo
  {author} {\bibfnamefont {J.}~\bibnamefont {Wilhelm}}, \bibinfo {author}
  {\bibfnamefont {J.}~\bibnamefont {Repp}}, \bibinfo {author} {\bibfnamefont
  {R.}~\bibnamefont {Huber}},\ and\ \bibinfo {author} {\bibfnamefont {Y.~A.}\
  \bibnamefont {Gerasimenko}},\ }\bibfield  {title} {\bibinfo {title}
  {Ultrafast atomic-scale scanning tunnelling spectroscopy of a single vacancy
  in a monolayer crystal},\ }\href {https://doi.org/10.1038/s41566-024-01390-6}
  {\bibfield  {journal} {\bibinfo  {journal} {Nature Photonics}\ }\textbf
  {\bibinfo {volume} {18}},\ \bibinfo {pages} {595} (\bibinfo {year}
  {2024})}\BibitemShut {NoStop}%
\bibitem [{\citenamefont {Xu}\ and\ \citenamefont {Zong}(2025)}]{Xu2025}%
  \BibitemOpen
  \bibfield  {author} {\bibinfo {author} {\bibfnamefont {C.}~\bibnamefont
  {Xu}}\ and\ \bibinfo {author} {\bibfnamefont {A.}~\bibnamefont {Zong}},\
  }\bibfield  {title} {\bibinfo {title} {Time-domain study of coupled
  collective excitations in quantum materials},\ }\href
  {https://doi.org/10.1038/s41535-025-00726-x} {\bibfield  {journal} {\bibinfo
  {journal} {npj Quantum Materials}\ }\textbf {\bibinfo {volume} {10}},\
  \bibinfo {pages} {21} (\bibinfo {year} {2025})}\BibitemShut {NoStop}%
\bibitem [{\citenamefont {Fausti}\ \emph {et~al.}(2011)\citenamefont {Fausti},
  \citenamefont {Tobey}, \citenamefont {Dean}, \citenamefont {Kaiser},
  \citenamefont {Dienst}, \citenamefont {Hoffmann}, \citenamefont {Pyon},
  \citenamefont {Takayama}, \citenamefont {Takagi},\ and\ \citenamefont
  {Cavalleri}}]{doi:10.1126/science.1197294}%
  \BibitemOpen
  \bibfield  {author} {\bibinfo {author} {\bibfnamefont {D.}~\bibnamefont
  {Fausti}}, \bibinfo {author} {\bibfnamefont {R.~I.}\ \bibnamefont {Tobey}},
  \bibinfo {author} {\bibfnamefont {N.}~\bibnamefont {Dean}}, \bibinfo {author}
  {\bibfnamefont {S.}~\bibnamefont {Kaiser}}, \bibinfo {author} {\bibfnamefont
  {A.}~\bibnamefont {Dienst}}, \bibinfo {author} {\bibfnamefont {M.~C.}\
  \bibnamefont {Hoffmann}}, \bibinfo {author} {\bibfnamefont {S.}~\bibnamefont
  {Pyon}}, \bibinfo {author} {\bibfnamefont {T.}~\bibnamefont {Takayama}},
  \bibinfo {author} {\bibfnamefont {H.}~\bibnamefont {Takagi}},\ and\ \bibinfo
  {author} {\bibfnamefont {A.}~\bibnamefont {Cavalleri}},\ }\bibfield  {title}
  {\bibinfo {title} {Light-induced superconductivity in a stripe-ordered
  cuprate},\ }\href {https://doi.org/10.1126/science.1197294} {\bibfield
  {journal} {\bibinfo  {journal} {Science}\ }\textbf {\bibinfo {volume}
  {331}},\ \bibinfo {pages} {189} (\bibinfo {year} {2011})}\BibitemShut
  {NoStop}%
\bibitem [{\citenamefont {Fava}\ \emph {et~al.}(2024)\citenamefont {Fava},
  \citenamefont {De~Vecchi}, \citenamefont {Jotzu}, \citenamefont {Buzzi},
  \citenamefont {Gebert}, \citenamefont {Liu}, \citenamefont {Keimer},\ and\
  \citenamefont {Cavalleri}}]{Fava2024}%
  \BibitemOpen
  \bibfield  {author} {\bibinfo {author} {\bibfnamefont {S.}~\bibnamefont
  {Fava}}, \bibinfo {author} {\bibfnamefont {G.}~\bibnamefont {De~Vecchi}},
  \bibinfo {author} {\bibfnamefont {G.}~\bibnamefont {Jotzu}}, \bibinfo
  {author} {\bibfnamefont {M.}~\bibnamefont {Buzzi}}, \bibinfo {author}
  {\bibfnamefont {T.}~\bibnamefont {Gebert}}, \bibinfo {author} {\bibfnamefont
  {Y.}~\bibnamefont {Liu}}, \bibinfo {author} {\bibfnamefont {B.}~\bibnamefont
  {Keimer}},\ and\ \bibinfo {author} {\bibfnamefont {A.}~\bibnamefont
  {Cavalleri}},\ }\bibfield  {title} {\bibinfo {title} {Magnetic field
  expulsion in optically driven {YB}a$_2${C}u$_3${O}$_{6.48}$},\ }\href
  {https://doi.org/10.1038/s41586-024-07635-2} {\bibfield  {journal} {\bibinfo
  {journal} {Nature}\ }\textbf {\bibinfo {volume} {632}},\ \bibinfo {pages}
  {75} (\bibinfo {year} {2024})}\BibitemShut {NoStop}%
\bibitem [{\citenamefont {Fechner}\ \emph {et~al.}(2018)\citenamefont
  {Fechner}, \citenamefont {Sukhov}, \citenamefont {Chotorlishvili},
  \citenamefont {Kenel}, \citenamefont {Berakdar},\ and\ \citenamefont
  {Spaldin}}]{PhysRevMaterials.2.064401}%
  \BibitemOpen
  \bibfield  {author} {\bibinfo {author} {\bibfnamefont {M.}~\bibnamefont
  {Fechner}}, \bibinfo {author} {\bibfnamefont {A.}~\bibnamefont {Sukhov}},
  \bibinfo {author} {\bibfnamefont {L.}~\bibnamefont {Chotorlishvili}},
  \bibinfo {author} {\bibfnamefont {C.}~\bibnamefont {Kenel}}, \bibinfo
  {author} {\bibfnamefont {J.}~\bibnamefont {Berakdar}},\ and\ \bibinfo
  {author} {\bibfnamefont {N.~A.}\ \bibnamefont {Spaldin}},\ }\bibfield
  {title} {\bibinfo {title} {Magnetophononics: {U}ltrafast spin control through
  the lattice},\ }\href {https://doi.org/10.1103/PhysRevMaterials.2.064401}
  {\bibfield  {journal} {\bibinfo  {journal} {Phys. Rev. Mater.}\ }\textbf
  {\bibinfo {volume} {2}},\ \bibinfo {pages} {064401} (\bibinfo {year}
  {2018})}\BibitemShut {NoStop}%
\bibitem [{\citenamefont {Giorgianni}\ \emph {et~al.}(2023)\citenamefont
  {Giorgianni}, \citenamefont {Wehinger}, \citenamefont {Allenspach},
  \citenamefont {Colonna}, \citenamefont {Vicario}, \citenamefont {Puphal},
  \citenamefont {Pomjakushina}, \citenamefont {Normand},\ and\ \citenamefont
  {R\"uegg}}]{PhysRevB.107.184440}%
  \BibitemOpen
  \bibfield  {author} {\bibinfo {author} {\bibfnamefont {F.}~\bibnamefont
  {Giorgianni}}, \bibinfo {author} {\bibfnamefont {B.}~\bibnamefont
  {Wehinger}}, \bibinfo {author} {\bibfnamefont {S.}~\bibnamefont
  {Allenspach}}, \bibinfo {author} {\bibfnamefont {N.}~\bibnamefont {Colonna}},
  \bibinfo {author} {\bibfnamefont {C.}~\bibnamefont {Vicario}}, \bibinfo
  {author} {\bibfnamefont {P.}~\bibnamefont {Puphal}}, \bibinfo {author}
  {\bibfnamefont {E.}~\bibnamefont {Pomjakushina}}, \bibinfo {author}
  {\bibfnamefont {B.}~\bibnamefont {Normand}},\ and\ \bibinfo {author}
  {\bibfnamefont {C.}~\bibnamefont {R\"uegg}},\ }\bibfield  {title} {\bibinfo
  {title} {Ultrafast frustration breaking and magnetophononic driving of
  singlet excitations in a quantum magnet},\ }\href
  {https://doi.org/10.1103/PhysRevB.107.184440} {\bibfield  {journal} {\bibinfo
   {journal} {Phys. Rev. B}\ }\textbf {\bibinfo {volume} {107}},\ \bibinfo
  {pages} {184440} (\bibinfo {year} {2023})}\BibitemShut {NoStop}%
\bibitem [{\citenamefont {Yarmohammadi}\ \emph {et~al.}(2021)\citenamefont
  {Yarmohammadi}, \citenamefont {Meyer}, \citenamefont {Fauseweh},
  \citenamefont {Normand},\ and\ \citenamefont {Uhrig}}]{PhysRevB.103.045132}%
  \BibitemOpen
  \bibfield  {author} {\bibinfo {author} {\bibfnamefont {M.}~\bibnamefont
  {Yarmohammadi}}, \bibinfo {author} {\bibfnamefont {C.}~\bibnamefont {Meyer}},
  \bibinfo {author} {\bibfnamefont {B.}~\bibnamefont {Fauseweh}}, \bibinfo
  {author} {\bibfnamefont {B.}~\bibnamefont {Normand}},\ and\ \bibinfo {author}
  {\bibfnamefont {G.~S.}\ \bibnamefont {Uhrig}},\ }\bibfield  {title} {\bibinfo
  {title} {Dynamical properties of a driven dissipative dimerized
  ${S}=\frac{1}{2}$ chain},\ }\href
  {https://doi.org/10.1103/PhysRevB.103.045132} {\bibfield  {journal} {\bibinfo
   {journal} {Phys. Rev. B}\ }\textbf {\bibinfo {volume} {103}},\ \bibinfo
  {pages} {045132} (\bibinfo {year} {2021})}\BibitemShut {NoStop}%
\bibitem [{\citenamefont {Walowski}\ and\ \citenamefont
  {M\"unzenberg}(2016)}]{10.1063/1.4958846}%
  \BibitemOpen
  \bibfield  {author} {\bibinfo {author} {\bibfnamefont {J.}~\bibnamefont
  {Walowski}}\ and\ \bibinfo {author} {\bibfnamefont {M.}~\bibnamefont
  {M\"unzenberg}},\ }\bibfield  {title} {\bibinfo {title} {Perspective:
  {U}ltrafast magnetism and {TH}z spintronics},\ }\href
  {https://doi.org/10.1063/1.4958846} {\bibfield  {journal} {\bibinfo
  {journal} {Journal of Applied Physics}\ }\textbf {\bibinfo {volume} {120}},\
  \bibinfo {pages} {140901} (\bibinfo {year} {2016})}\BibitemShut {NoStop}%
\bibitem [{\citenamefont {Luo}\ \emph {et~al.}(2023)\citenamefont {Luo},
  \citenamefont {Lin}, \citenamefont {Zhang}, \citenamefont {Chen},
  \citenamefont {Blackert}, \citenamefont {Xu}, \citenamefont {Yakobson},\ and\
  \citenamefont {Zhu}}]{doi:10.1126/science.adi9601}%
  \BibitemOpen
  \bibfield  {author} {\bibinfo {author} {\bibfnamefont {J.}~\bibnamefont
  {Luo}}, \bibinfo {author} {\bibfnamefont {T.}~\bibnamefont {Lin}}, \bibinfo
  {author} {\bibfnamefont {J.}~\bibnamefont {Zhang}}, \bibinfo {author}
  {\bibfnamefont {X.}~\bibnamefont {Chen}}, \bibinfo {author} {\bibfnamefont
  {E.~R.}\ \bibnamefont {Blackert}}, \bibinfo {author} {\bibfnamefont
  {R.}~\bibnamefont {Xu}}, \bibinfo {author} {\bibfnamefont {B.~I.}\
  \bibnamefont {Yakobson}},\ and\ \bibinfo {author} {\bibfnamefont
  {H.}~\bibnamefont {Zhu}},\ }\bibfield  {title} {\bibinfo {title} {Large
  effective magnetic fields from chiral phonons in rare-earth halides},\ }\href
  {https://doi.org/10.1126/science.adi9601} {\bibfield  {journal} {\bibinfo
  {journal} {Science}\ }\textbf {\bibinfo {volume} {382}},\ \bibinfo {pages}
  {698} (\bibinfo {year} {2023})}\BibitemShut {NoStop}%
\bibitem [{\citenamefont {Merlin}(2025)}]{10.1093/pnasnexus/pgaf002}%
  \BibitemOpen
  \bibfield  {author} {\bibinfo {author} {\bibfnamefont {R.}~\bibnamefont
  {Merlin}},\ }\bibfield  {title} {\bibinfo {title} {Magnetophononics and the
  chiral phonon misnomer},\ }\href {https://doi.org/10.1093/pnasnexus/pgaf002}
  {\bibfield  {journal} {\bibinfo  {journal} {PNAS Nexus}\ }\textbf {\bibinfo
  {volume} {4}},\ \bibinfo {pages} {pgaf002} (\bibinfo {year}
  {2025})}\BibitemShut {NoStop}%
\bibitem [{\citenamefont {Yarmohammadi}\ \emph {et~al.}(2023)\citenamefont
  {Yarmohammadi}, \citenamefont {Krebs}, \citenamefont {Uhrig},\ and\
  \citenamefont {Normand}}]{PhysRevB.107.174415}%
  \BibitemOpen
  \bibfield  {author} {\bibinfo {author} {\bibfnamefont {M.}~\bibnamefont
  {Yarmohammadi}}, \bibinfo {author} {\bibfnamefont {M.}~\bibnamefont {Krebs}},
  \bibinfo {author} {\bibfnamefont {G.~S.}\ \bibnamefont {Uhrig}},\ and\
  \bibinfo {author} {\bibfnamefont {B.}~\bibnamefont {Normand}},\ }\bibfield
  {title} {\bibinfo {title} {Strong-coupling magnetophononics: Self-blocking,
  phonon-bitriplons, and spin-band engineering},\ }\href
  {https://doi.org/10.1103/PhysRevB.107.174415} {\bibfield  {journal} {\bibinfo
   {journal} {Phys. Rev. B}\ }\textbf {\bibinfo {volume} {107}},\ \bibinfo
  {pages} {174415} (\bibinfo {year} {2023})}\BibitemShut {NoStop}%
\bibitem [{\citenamefont {Allafi}\ \emph {et~al.}(2024)\citenamefont {Allafi},
  \citenamefont {Kolodrubetz}, \citenamefont {Bukov}, \citenamefont
  {Oganesyan},\ and\ \citenamefont {Yarmohammadi}}]{PhysRevB.110.064420}%
  \BibitemOpen
  \bibfield  {author} {\bibinfo {author} {\bibfnamefont {N.~M.}\ \bibnamefont
  {Allafi}}, \bibinfo {author} {\bibfnamefont {M.~H.}\ \bibnamefont
  {Kolodrubetz}}, \bibinfo {author} {\bibfnamefont {M.}~\bibnamefont {Bukov}},
  \bibinfo {author} {\bibfnamefont {V.}~\bibnamefont {Oganesyan}},\ and\
  \bibinfo {author} {\bibfnamefont {M.}~\bibnamefont {Yarmohammadi}},\
  }\bibfield  {title} {\bibinfo {title} {Spin high harmonic generation through
  terahertz laser-driven phonons},\ }\href
  {https://doi.org/10.1103/PhysRevB.110.064420} {\bibfield  {journal} {\bibinfo
   {journal} {Phys. Rev. B}\ }\textbf {\bibinfo {volume} {110}},\ \bibinfo
  {pages} {064420} (\bibinfo {year} {2024})}\BibitemShut {NoStop}%
\bibitem [{\citenamefont {Yarmohammadi}\ and\ \citenamefont
  {Kolodrubetz}(2024)}]{PhysRevB.110.134442}%
  \BibitemOpen
  \bibfield  {author} {\bibinfo {author} {\bibfnamefont {M.}~\bibnamefont
  {Yarmohammadi}}\ and\ \bibinfo {author} {\bibfnamefont {M.~H.}\ \bibnamefont
  {Kolodrubetz}},\ }\bibfield  {title} {\bibinfo {title} {Terahertz
  high-harmonic generation in gapped antiferromagnetic chains},\ }\href
  {https://doi.org/10.1103/PhysRevB.110.134442} {\bibfield  {journal} {\bibinfo
   {journal} {Phys. Rev. B}\ }\textbf {\bibinfo {volume} {110}},\ \bibinfo
  {pages} {134442} (\bibinfo {year} {2024})}\BibitemShut {NoStop}%
\bibitem [{\citenamefont {Demazure}\ \emph {et~al.}(2025)\citenamefont
  {Demazure}, \citenamefont {Krebs}, \citenamefont {Uhrig},\ and\ \citenamefont
  {Normand}}]{4ddn-y88c}%
  \BibitemOpen
  \bibfield  {author} {\bibinfo {author} {\bibfnamefont {B.}~\bibnamefont
  {Demazure}}, \bibinfo {author} {\bibfnamefont {M.}~\bibnamefont {Krebs}},
  \bibinfo {author} {\bibfnamefont {G.~S.}\ \bibnamefont {Uhrig}},\ and\
  \bibinfo {author} {\bibfnamefont {B.}~\bibnamefont {Normand}},\ }\bibfield
  {title} {\bibinfo {title} {Pulsed magnetophononics in gapped quantum
  magnets},\ }\href {https://doi.org/10.1103/4ddn-y88c} {\bibfield  {journal}
  {\bibinfo  {journal} {Phys. Rev. B}\ }\textbf {\bibinfo {volume} {112}},\
  \bibinfo {pages} {075112} (\bibinfo {year} {2025})}\BibitemShut {NoStop}%
\bibitem [{\citenamefont {Yarmohammadi}(2025)}]{c3tb-h5hv}%
  \BibitemOpen
  \bibfield  {author} {\bibinfo {author} {\bibfnamefont {M.}~\bibnamefont
  {Yarmohammadi}},\ }\bibfield  {title} {\bibinfo {title} {Dynamical phase
  transition in a strongly hybridized phonon-triplon chain},\ }\href
  {https://doi.org/10.1103/c3tb-h5hv} {\bibfield  {journal} {\bibinfo
  {journal} {Phys. Rev. B}\ }\textbf {\bibinfo {volume} {112}},\ \bibinfo
  {pages} {144432} (\bibinfo {year} {2025})}\BibitemShut {NoStop}%
\bibitem [{\citenamefont {Kahana}\ \emph {et~al.}(2024)\citenamefont {Kahana},
  \citenamefont {Lopez},\ and\ \citenamefont
  {Juraschek}}]{doi:10.1126/sciadv.ado0722}%
  \BibitemOpen
  \bibfield  {author} {\bibinfo {author} {\bibfnamefont {T.}~\bibnamefont
  {Kahana}}, \bibinfo {author} {\bibfnamefont {D.~A.~B.}\ \bibnamefont
  {Lopez}},\ and\ \bibinfo {author} {\bibfnamefont {D.~M.}\ \bibnamefont
  {Juraschek}},\ }\bibfield  {title} {\bibinfo {title} {Light-induced
  magnetization from magnonic rectification},\ }\href
  {https://doi.org/10.1126/sciadv.ado0722} {\bibfield  {journal} {\bibinfo
  {journal} {Science Advances}\ }\textbf {\bibinfo {volume} {10}},\ \bibinfo
  {pages} {eado0722} (\bibinfo {year} {2024})}\BibitemShut {NoStop}%
\bibitem [{\citenamefont {Knetter}\ and\ \citenamefont
  {Uhrig}(2001)}]{PhysRevB.63.094401}%
  \BibitemOpen
  \bibfield  {author} {\bibinfo {author} {\bibfnamefont {C.}~\bibnamefont
  {Knetter}}\ and\ \bibinfo {author} {\bibfnamefont {G.~S.}\ \bibnamefont
  {Uhrig}},\ }\bibfield  {title} {\bibinfo {title} {Triplet dispersion in
  {C}u{G}e{O}$_{3}:$ {P}erturbative analysis},\ }\href
  {https://doi.org/10.1103/PhysRevB.63.094401} {\bibfield  {journal} {\bibinfo
  {journal} {Phys. Rev. B}\ }\textbf {\bibinfo {volume} {63}},\ \bibinfo
  {pages} {094401} (\bibinfo {year} {2001})}\BibitemShut {NoStop}%
\bibitem [{\citenamefont {Chen}\ \emph {et~al.}(2021)\citenamefont {Chen},
  \citenamefont {Sato}, \citenamefont {Tang}, \citenamefont {Shiomi},
  \citenamefont {Oyanagi}, \citenamefont {Masuda}, \citenamefont {Nambu},
  \citenamefont {Fujita},\ and\ \citenamefont {Saitoh}}]{Chen2021}%
  \BibitemOpen
  \bibfield  {author} {\bibinfo {author} {\bibfnamefont {Y.}~\bibnamefont
  {Chen}}, \bibinfo {author} {\bibfnamefont {M.}~\bibnamefont {Sato}}, \bibinfo
  {author} {\bibfnamefont {Y.}~\bibnamefont {Tang}}, \bibinfo {author}
  {\bibfnamefont {Y.}~\bibnamefont {Shiomi}}, \bibinfo {author} {\bibfnamefont
  {K.}~\bibnamefont {Oyanagi}}, \bibinfo {author} {\bibfnamefont
  {T.}~\bibnamefont {Masuda}}, \bibinfo {author} {\bibfnamefont
  {Y.}~\bibnamefont {Nambu}}, \bibinfo {author} {\bibfnamefont
  {M.}~\bibnamefont {Fujita}},\ and\ \bibinfo {author} {\bibfnamefont
  {E.}~\bibnamefont {Saitoh}},\ }\bibfield  {title} {\bibinfo {title} {Triplon
  current generation in solids},\ }\href
  {https://doi.org/10.1038/s41467-021-25494-7} {\bibfield  {journal} {\bibinfo
  {journal} {Nature Communications}\ }\textbf {\bibinfo {volume} {12}},\
  \bibinfo {pages} {5199} (\bibinfo {year} {2021})}\BibitemShut {NoStop}%
\bibitem [{\citenamefont {Park}\ \emph {et~al.}(2025)\citenamefont {Park},
  \citenamefont {Xiao}, \citenamefont {Górnicka}, \citenamefont {May},
  \citenamefont {Yan}, \citenamefont {Kajimoto}, \citenamefont {Nakamura},
  \citenamefont {Stone}, \citenamefont {Halász},\ and\ \citenamefont
  {Christianson}}]{park2025weaklyinteractingspinonstightly}%
  \BibitemOpen
  \bibfield  {author} {\bibinfo {author} {\bibfnamefont {P.}~\bibnamefont
  {Park}}, \bibinfo {author} {\bibfnamefont {B.}~\bibnamefont {Xiao}}, \bibinfo
  {author} {\bibfnamefont {K.}~\bibnamefont {Górnicka}}, \bibinfo {author}
  {\bibfnamefont {A.~F.}\ \bibnamefont {May}}, \bibinfo {author} {\bibfnamefont
  {J.}~\bibnamefont {Yan}}, \bibinfo {author} {\bibfnamefont {R.}~\bibnamefont
  {Kajimoto}}, \bibinfo {author} {\bibfnamefont {M.}~\bibnamefont {Nakamura}},
  \bibinfo {author} {\bibfnamefont {M.~B.}\ \bibnamefont {Stone}}, \bibinfo
  {author} {\bibfnamefont {G.~B.}\ \bibnamefont {Halász}},\ and\ \bibinfo
  {author} {\bibfnamefont {A.~D.}\ \bibnamefont {Christianson}},\ }\bibfield
  {title} {\bibinfo {title} {From weakly interacting spinons to tightly bound
  triplons in the frustrated quantum spin-{P}eierls chain},\ }\href
  {https://arxiv.org/abs/2507.19412} {\  (\bibinfo {year} {2025})},\ \Eprint
  {https://arxiv.org/abs/2507.19412} {arXiv:2507.19412 [cond-mat.str-el]}
  \BibitemShut {NoStop}%
\bibitem [{\citenamefont {{van Loosdrecht}}\ \emph {et~al.}(1997)\citenamefont
  {{van Loosdrecht}}, \citenamefont {Boucher}, \citenamefont {Huant},
  \citenamefont {Martinez}, \citenamefont {Dhalenne},\ and\ \citenamefont
  {Revcolevschi}}]{VANLOOSDRECHT19971017}%
  \BibitemOpen
  \bibfield  {author} {\bibinfo {author} {\bibfnamefont {P.}~\bibnamefont {{van
  Loosdrecht}}}, \bibinfo {author} {\bibfnamefont {J.}~\bibnamefont {Boucher}},
  \bibinfo {author} {\bibfnamefont {S.}~\bibnamefont {Huant}}, \bibinfo
  {author} {\bibfnamefont {G.}~\bibnamefont {Martinez}}, \bibinfo {author}
  {\bibfnamefont {G.}~\bibnamefont {Dhalenne}},\ and\ \bibinfo {author}
  {\bibfnamefont {A.}~\bibnamefont {Revcolevschi}},\ }\bibfield  {title}
  {\bibinfo {title} {Spins and phonons in the spin-{P}eierls compound
  {C}u{G}e{O}$_3$},\ }\href
  {https://doi.org/https://doi.org/10.1016/S0921-4526(96)00793-4} {\bibfield
  {journal} {\bibinfo  {journal} {Physica B: Condensed Matter}\ }\textbf
  {\bibinfo {volume} {230-232}},\ \bibinfo {pages} {1017} (\bibinfo {year}
  {1997})},\ \bibinfo {note} {proceedings of the International Conference on
  Strongly Correlated Electron Systems}\BibitemShut {NoStop}%
\bibitem [{\citenamefont {Büchner}\ \emph {et~al.}(1999)\citenamefont
  {Büchner}, \citenamefont {Fehske}, \citenamefont {Kampf},\ and\
  \citenamefont {Wellein}}]{BUCHNER1999956}%
  \BibitemOpen
  \bibfield  {author} {\bibinfo {author} {\bibfnamefont {B.}~\bibnamefont
  {Büchner}}, \bibinfo {author} {\bibfnamefont {H.}~\bibnamefont {Fehske}},
  \bibinfo {author} {\bibfnamefont {A.}~\bibnamefont {Kampf}},\ and\ \bibinfo
  {author} {\bibfnamefont {G.}~\bibnamefont {Wellein}},\ }\bibfield  {title}
  {\bibinfo {title} {Lattice dimerization in the spin–{P}eierls compound
  {C}u{G}e{O}$_3$},\ }\href
  {https://doi.org/https://doi.org/10.1016/S0921-4526(98)00720-0} {\bibfield
  {journal} {\bibinfo  {journal} {Physica B: Condensed Matter}\ }\textbf
  {\bibinfo {volume} {259-261}},\ \bibinfo {pages} {956} (\bibinfo {year}
  {1999})}\BibitemShut {NoStop}%
\bibitem [{\citenamefont {Yuasa}\ \emph {et~al.}(2008)\citenamefont {Yuasa},
  \citenamefont {Nakajima}, \citenamefont {Yamanouchi}, \citenamefont {Ueda},\
  and\ \citenamefont {Suemoto}}]{YUASA20081087}%
  \BibitemOpen
  \bibfield  {author} {\bibinfo {author} {\bibfnamefont {Y.}~\bibnamefont
  {Yuasa}}, \bibinfo {author} {\bibfnamefont {M.}~\bibnamefont {Nakajima}},
  \bibinfo {author} {\bibfnamefont {T.}~\bibnamefont {Yamanouchi}}, \bibinfo
  {author} {\bibfnamefont {Y.}~\bibnamefont {Ueda}},\ and\ \bibinfo {author}
  {\bibfnamefont {T.}~\bibnamefont {Suemoto}},\ }\bibfield  {title} {\bibinfo
  {title} {Ultrafast time-resolved spectroscopy of the spin-{P}eierls compound
  {C}u{G}e{O}$_3$},\ }\href
  {https://doi.org/https://doi.org/10.1016/j.jlumin.2007.11.005} {\bibfield
  {journal} {\bibinfo  {journal} {Journal of Luminescence}\ }\textbf {\bibinfo
  {volume} {128}},\ \bibinfo {pages} {1087} (\bibinfo {year} {2008})},\
  \bibinfo {note} {proceedings of the 16th International Conference on
  Dynamical Processes in Excited States of Solids}\BibitemShut {NoStop}%
\bibitem [{\citenamefont {R\"uckamp}\ \emph {et~al.}(2005)\citenamefont
  {R\"uckamp}, \citenamefont {Baier}, \citenamefont {Kriener}, \citenamefont
  {Haverkort}, \citenamefont {Lorenz}, \citenamefont {Uhrig}, \citenamefont
  {Jongen}, \citenamefont {M\"oller}, \citenamefont {Meyer},\ and\
  \citenamefont {Gr\"uninger}}]{PhysRevLett.95.097203}%
  \BibitemOpen
  \bibfield  {author} {\bibinfo {author} {\bibfnamefont {R.}~\bibnamefont
  {R\"uckamp}}, \bibinfo {author} {\bibfnamefont {J.}~\bibnamefont {Baier}},
  \bibinfo {author} {\bibfnamefont {M.}~\bibnamefont {Kriener}}, \bibinfo
  {author} {\bibfnamefont {M.~W.}\ \bibnamefont {Haverkort}}, \bibinfo {author}
  {\bibfnamefont {T.}~\bibnamefont {Lorenz}}, \bibinfo {author} {\bibfnamefont
  {G.~S.}\ \bibnamefont {Uhrig}}, \bibinfo {author} {\bibfnamefont
  {L.}~\bibnamefont {Jongen}}, \bibinfo {author} {\bibfnamefont
  {A.}~\bibnamefont {M\"oller}}, \bibinfo {author} {\bibfnamefont
  {G.}~\bibnamefont {Meyer}},\ and\ \bibinfo {author} {\bibfnamefont
  {M.}~\bibnamefont {Gr\"uninger}},\ }\bibfield  {title} {\bibinfo {title}
  {Zero-field incommensurate spin-{P}eierls phase with interchain frustration
  in {T}i{OC}l},\ }\href {https://doi.org/10.1103/PhysRevLett.95.097203}
  {\bibfield  {journal} {\bibinfo  {journal} {Phys. Rev. Lett.}\ }\textbf
  {\bibinfo {volume} {95}},\ \bibinfo {pages} {097203} (\bibinfo {year}
  {2005})}\BibitemShut {NoStop}%
\bibitem [{\citenamefont {Shaz}\ \emph {et~al.}(2005)\citenamefont {Shaz},
  \citenamefont {van Smaalen}, \citenamefont {Palatinus}, \citenamefont
  {Hoinkis}, \citenamefont {Klemm}, \citenamefont {Horn},\ and\ \citenamefont
  {Claessen}}]{PhysRevB.71.100405}%
  \BibitemOpen
  \bibfield  {author} {\bibinfo {author} {\bibfnamefont {M.}~\bibnamefont
  {Shaz}}, \bibinfo {author} {\bibfnamefont {S.}~\bibnamefont {van Smaalen}},
  \bibinfo {author} {\bibfnamefont {L.}~\bibnamefont {Palatinus}}, \bibinfo
  {author} {\bibfnamefont {M.}~\bibnamefont {Hoinkis}}, \bibinfo {author}
  {\bibfnamefont {M.}~\bibnamefont {Klemm}}, \bibinfo {author} {\bibfnamefont
  {S.}~\bibnamefont {Horn}},\ and\ \bibinfo {author} {\bibfnamefont
  {R.}~\bibnamefont {Claessen}},\ }\bibfield  {title} {\bibinfo {title}
  {Spin-{P}eierls transition in {T}i{OC}l},\ }\href
  {https://doi.org/10.1103/PhysRevB.71.100405} {\bibfield  {journal} {\bibinfo
  {journal} {Phys. Rev. B}\ }\textbf {\bibinfo {volume} {71}},\ \bibinfo
  {pages} {100405(R)} (\bibinfo {year} {2005})}\BibitemShut {NoStop}%
\bibitem [{\citenamefont {Glawion}\ \emph {et~al.}(2011)\citenamefont
  {Glawion}, \citenamefont {Heidler}, \citenamefont {Haverkort}, \citenamefont
  {Duda}, \citenamefont {Schmitt}, \citenamefont {Strocov}, \citenamefont
  {Monney}, \citenamefont {Zhou}, \citenamefont {Ruff}, \citenamefont {Sing},\
  and\ \citenamefont {Claessen}}]{PhysRevLett.107.107402}%
  \BibitemOpen
  \bibfield  {author} {\bibinfo {author} {\bibfnamefont {S.}~\bibnamefont
  {Glawion}}, \bibinfo {author} {\bibfnamefont {J.}~\bibnamefont {Heidler}},
  \bibinfo {author} {\bibfnamefont {M.~W.}\ \bibnamefont {Haverkort}}, \bibinfo
  {author} {\bibfnamefont {L.~C.}\ \bibnamefont {Duda}}, \bibinfo {author}
  {\bibfnamefont {T.}~\bibnamefont {Schmitt}}, \bibinfo {author} {\bibfnamefont
  {V.~N.}\ \bibnamefont {Strocov}}, \bibinfo {author} {\bibfnamefont
  {C.}~\bibnamefont {Monney}}, \bibinfo {author} {\bibfnamefont {K.~J.}\
  \bibnamefont {Zhou}}, \bibinfo {author} {\bibfnamefont {A.}~\bibnamefont
  {Ruff}}, \bibinfo {author} {\bibfnamefont {M.}~\bibnamefont {Sing}},\ and\
  \bibinfo {author} {\bibfnamefont {R.}~\bibnamefont {Claessen}},\ }\bibfield
  {title} {\bibinfo {title} {Two-spinon and orbital excitations of the
  spin-{P}eierls system {T}i{OC}l},\ }\href
  {https://doi.org/10.1103/PhysRevLett.107.107402} {\bibfield  {journal}
  {\bibinfo  {journal} {Phys. Rev. Lett.}\ }\textbf {\bibinfo {volume} {107}},\
  \bibinfo {pages} {107402} (\bibinfo {year} {2011})}\BibitemShut {NoStop}%
\bibitem [{\citenamefont {Ikeda}\ and\ \citenamefont {Sato}(2020)}]{Ikeda2020}%
  \BibitemOpen
  \bibfield  {author} {\bibinfo {author} {\bibfnamefont {T.~N.}\ \bibnamefont
  {Ikeda}}\ and\ \bibinfo {author} {\bibfnamefont {M.}~\bibnamefont {Sato}},\
  }\bibfield  {title} {\bibinfo {title} {General description for nonequilibrium
  steady states in periodically driven dissipative quantum systems},\ }\href
  {https://doi.org/10.1126/sciadv.abb4019} {\bibfield  {journal} {\bibinfo
  {journal} {Science Advances}\ }\textbf {\bibinfo {volume} {6}},\ \bibinfo
  {pages} {eabb4019} (\bibinfo {year} {2020})}\BibitemShut {NoStop}%
\bibitem [{\citenamefont {Sieberer}\ \emph {et~al.}(2025)\citenamefont
  {Sieberer}, \citenamefont {Buchhold}, \citenamefont {Marino},\ and\
  \citenamefont {Diehl}}]{RevModPhys.97.025004}%
  \BibitemOpen
  \bibfield  {author} {\bibinfo {author} {\bibfnamefont {L.~M.}\ \bibnamefont
  {Sieberer}}, \bibinfo {author} {\bibfnamefont {M.}~\bibnamefont {Buchhold}},
  \bibinfo {author} {\bibfnamefont {J.}~\bibnamefont {Marino}},\ and\ \bibinfo
  {author} {\bibfnamefont {S.}~\bibnamefont {Diehl}},\ }\bibfield  {title}
  {\bibinfo {title} {Universality in driven open quantum matter},\ }\href
  {https://doi.org/10.1103/RevModPhys.97.025004} {\bibfield  {journal}
  {\bibinfo  {journal} {Rev. Mod. Phys.}\ }\textbf {\bibinfo {volume} {97}},\
  \bibinfo {pages} {025004} (\bibinfo {year} {2025})}\BibitemShut {NoStop}%
\bibitem [{\citenamefont {Werner}\ \emph {et~al.}(1999)\citenamefont {Werner},
  \citenamefont {Gros},\ and\ \citenamefont {Braden}}]{PhysRevB.59.14356}%
  \BibitemOpen
  \bibfield  {author} {\bibinfo {author} {\bibfnamefont {R.}~\bibnamefont
  {Werner}}, \bibinfo {author} {\bibfnamefont {C.}~\bibnamefont {Gros}},\ and\
  \bibinfo {author} {\bibfnamefont {M.}~\bibnamefont {Braden}},\ }\bibfield
  {title} {\bibinfo {title} {Microscopic spin-phonon coupling constants in
  {C}u{G}e{O}$_{3}$},\ }\href {https://doi.org/10.1103/PhysRevB.59.14356}
  {\bibfield  {journal} {\bibinfo  {journal} {Phys. Rev. B}\ }\textbf {\bibinfo
  {volume} {59}},\ \bibinfo {pages} {14356} (\bibinfo {year}
  {1999})}\BibitemShut {NoStop}%
\bibitem [{\citenamefont {Spitz}\ \emph {et~al.}(2025)\citenamefont {Spitz},
  \citenamefont {Razpopov}, \citenamefont {Biswas}, \citenamefont {Lane},
  \citenamefont {Nikitin}, \citenamefont {Iida}, \citenamefont {Kajimoto},
  \citenamefont {Fujita}, \citenamefont {Arai}, \citenamefont {Mourigal},
  \citenamefont {R\"uegg}, \citenamefont {Valent\'{\i}},\ and\ \citenamefont
  {Normand}}]{jz36-8kz9}%
  \BibitemOpen
  \bibfield  {author} {\bibinfo {author} {\bibfnamefont {L.}~\bibnamefont
  {Spitz}}, \bibinfo {author} {\bibfnamefont {A.}~\bibnamefont {Razpopov}},
  \bibinfo {author} {\bibfnamefont {S.}~\bibnamefont {Biswas}}, \bibinfo
  {author} {\bibfnamefont {H.}~\bibnamefont {Lane}}, \bibinfo {author}
  {\bibfnamefont {S.~E.}\ \bibnamefont {Nikitin}}, \bibinfo {author}
  {\bibfnamefont {K.}~\bibnamefont {Iida}}, \bibinfo {author} {\bibfnamefont
  {R.}~\bibnamefont {Kajimoto}}, \bibinfo {author} {\bibfnamefont
  {M.}~\bibnamefont {Fujita}}, \bibinfo {author} {\bibfnamefont
  {M.}~\bibnamefont {Arai}}, \bibinfo {author} {\bibfnamefont {M.}~\bibnamefont
  {Mourigal}}, \bibinfo {author} {\bibfnamefont {C.}~\bibnamefont {R\"uegg}},
  \bibinfo {author} {\bibfnamefont {R.}~\bibnamefont {Valent\'{\i}}},\ and\
  \bibinfo {author} {\bibfnamefont {B.}~\bibnamefont {Normand}},\ }\bibfield
  {title} {\bibinfo {title} {Phonon spectrum in the spin-{P}eierls phase of
  {C}u{G}e{O}$_{3}$},\ }\href {https://doi.org/10.1103/jz36-8kz9} {\bibfield
  {journal} {\bibinfo  {journal} {Phys. Rev. B}\ }\textbf {\bibinfo {volume}
  {112}},\ \bibinfo {pages} {184302} (\bibinfo {year} {2025})}\BibitemShut
  {NoStop}%
\bibitem [{\citenamefont {Kuroe}\ \emph {et~al.}(1994)\citenamefont {Kuroe},
  \citenamefont {Sekine}, \citenamefont {Hase}, \citenamefont {Sasago},
  \citenamefont {Uchinokura}, \citenamefont {Kojima}, \citenamefont {Tanaka},\
  and\ \citenamefont {Shibuya}}]{PhysRevB.50.16468}%
  \BibitemOpen
  \bibfield  {author} {\bibinfo {author} {\bibfnamefont {H.}~\bibnamefont
  {Kuroe}}, \bibinfo {author} {\bibfnamefont {T.}~\bibnamefont {Sekine}},
  \bibinfo {author} {\bibfnamefont {M.}~\bibnamefont {Hase}}, \bibinfo {author}
  {\bibfnamefont {Y.}~\bibnamefont {Sasago}}, \bibinfo {author} {\bibfnamefont
  {K.}~\bibnamefont {Uchinokura}}, \bibinfo {author} {\bibfnamefont
  {H.}~\bibnamefont {Kojima}}, \bibinfo {author} {\bibfnamefont
  {I.}~\bibnamefont {Tanaka}},\ and\ \bibinfo {author} {\bibfnamefont
  {Y.}~\bibnamefont {Shibuya}},\ }\bibfield  {title} {\bibinfo {title}
  {{R}aman-scattering study of {C}u{G}e{O}$_{3}$ in the spin-{P}eierls phase},\
  }\href {https://doi.org/10.1103/PhysRevB.50.16468} {\bibfield  {journal}
  {\bibinfo  {journal} {Phys. Rev. B}\ }\textbf {\bibinfo {volume} {50}},\
  \bibinfo {pages} {16468} (\bibinfo {year} {1994})}\BibitemShut {NoStop}%
\bibitem [{\citenamefont {Popovi\ifmmode~\acute{c}\else \'{c}\fi{}}\ \emph
  {et~al.}(1995)\citenamefont {Popovi\ifmmode~\acute{c}\else \'{c}\fi{}},
  \citenamefont {Devi\ifmmode~\acute{c}\else \'{c}\fi{}}, \citenamefont
  {Popov}, \citenamefont {Dhalenne},\ and\ \citenamefont
  {Revcolevschi}}]{PhysRevB.52.4185}%
  \BibitemOpen
  \bibfield  {author} {\bibinfo {author} {\bibfnamefont {Z.~V.}\ \bibnamefont
  {Popovi\ifmmode~\acute{c}\else \'{c}\fi{}}}, \bibinfo {author} {\bibfnamefont
  {S.~D.}\ \bibnamefont {Devi\ifmmode~\acute{c}\else \'{c}\fi{}}}, \bibinfo
  {author} {\bibfnamefont {V.~N.}\ \bibnamefont {Popov}}, \bibinfo {author}
  {\bibfnamefont {G.}~\bibnamefont {Dhalenne}},\ and\ \bibinfo {author}
  {\bibfnamefont {A.}~\bibnamefont {Revcolevschi}},\ }\bibfield  {title}
  {\bibinfo {title} {Phonons in {C}u{G}e{O}$_{3}$ studied using polarized
  far-infrared and {R}aman-scattering spectroscopies},\ }\href
  {https://doi.org/10.1103/PhysRevB.52.4185} {\bibfield  {journal} {\bibinfo
  {journal} {Phys. Rev. B}\ }\textbf {\bibinfo {volume} {52}},\ \bibinfo
  {pages} {4185} (\bibinfo {year} {1995})}\BibitemShut {NoStop}%
\bibitem [{\citenamefont {Henstridge}\ \emph {et~al.}(2022)\citenamefont
  {Henstridge}, \citenamefont {F{\"o}rst}, \citenamefont {Rowe}, \citenamefont
  {Fechner},\ and\ \citenamefont {Cavalleri}}]{Henstridge2022}%
  \BibitemOpen
  \bibfield  {author} {\bibinfo {author} {\bibfnamefont {M.}~\bibnamefont
  {Henstridge}}, \bibinfo {author} {\bibfnamefont {M.}~\bibnamefont
  {F{\"o}rst}}, \bibinfo {author} {\bibfnamefont {E.}~\bibnamefont {Rowe}},
  \bibinfo {author} {\bibfnamefont {M.}~\bibnamefont {Fechner}},\ and\ \bibinfo
  {author} {\bibfnamefont {A.}~\bibnamefont {Cavalleri}},\ }\bibfield  {title}
  {\bibinfo {title} {Nonlocal nonlinear phononics},\ }\href
  {https://doi.org/10.1038/s41567-022-01512-3} {\bibfield  {journal} {\bibinfo
  {journal} {Nature Physics}\ }\textbf {\bibinfo {volume} {18}},\ \bibinfo
  {pages} {457} (\bibinfo {year} {2022})}\BibitemShut {NoStop}%
\bibitem [{\citenamefont {Juraschek}\ \emph {et~al.}(2021)\citenamefont
  {Juraschek}, \citenamefont {Neuman}, \citenamefont {Flick},\ and\
  \citenamefont {Narang}}]{PhysRevResearch.3.L032046}%
  \BibitemOpen
  \bibfield  {author} {\bibinfo {author} {\bibfnamefont {D.~M.}\ \bibnamefont
  {Juraschek}}, \bibinfo {author} {\bibfnamefont {T.~c.~v.}\ \bibnamefont
  {Neuman}}, \bibinfo {author} {\bibfnamefont {J.}~\bibnamefont {Flick}},\ and\
  \bibinfo {author} {\bibfnamefont {P.}~\bibnamefont {Narang}},\ }\bibfield
  {title} {\bibinfo {title} {Cavity control of nonlinear phononics},\ }\href
  {https://doi.org/10.1103/PhysRevResearch.3.L032046} {\bibfield  {journal}
  {\bibinfo  {journal} {Phys. Rev. Res.}\ }\textbf {\bibinfo {volume} {3}},\
  \bibinfo {pages} {L032046} (\bibinfo {year} {2021})}\BibitemShut {NoStop}%
\bibitem [{\citenamefont {Subedi}\ \emph {et~al.}(2014)\citenamefont {Subedi},
  \citenamefont {Cavalleri},\ and\ \citenamefont
  {Georges}}]{PhysRevB.89.220301}%
  \BibitemOpen
  \bibfield  {author} {\bibinfo {author} {\bibfnamefont {A.}~\bibnamefont
  {Subedi}}, \bibinfo {author} {\bibfnamefont {A.}~\bibnamefont {Cavalleri}},\
  and\ \bibinfo {author} {\bibfnamefont {A.}~\bibnamefont {Georges}},\
  }\bibfield  {title} {\bibinfo {title} {Theory of nonlinear phononics for
  coherent light control of solids},\ }\href
  {https://doi.org/10.1103/PhysRevB.89.220301} {\bibfield  {journal} {\bibinfo
  {journal} {Phys. Rev. B}\ }\textbf {\bibinfo {volume} {89}},\ \bibinfo
  {pages} {220301(R)} (\bibinfo {year} {2014})}\BibitemShut {NoStop}%
\bibitem [{\citenamefont {Juraschek}\ \emph {et~al.}(2017)\citenamefont
  {Juraschek}, \citenamefont {Fechner},\ and\ \citenamefont
  {Spaldin}}]{PhysRevLett.118.054101}%
  \BibitemOpen
  \bibfield  {author} {\bibinfo {author} {\bibfnamefont {D.~M.}\ \bibnamefont
  {Juraschek}}, \bibinfo {author} {\bibfnamefont {M.}~\bibnamefont {Fechner}},\
  and\ \bibinfo {author} {\bibfnamefont {N.~A.}\ \bibnamefont {Spaldin}},\
  }\bibfield  {title} {\bibinfo {title} {Ultrafast structure switching through
  nonlinear phononics},\ }\href
  {https://doi.org/10.1103/PhysRevLett.118.054101} {\bibfield  {journal}
  {\bibinfo  {journal} {Phys. Rev. Lett.}\ }\textbf {\bibinfo {volume} {118}},\
  \bibinfo {pages} {054101} (\bibinfo {year} {2017})}\BibitemShut {NoStop}%
\bibitem [{\citenamefont {Rini}\ \emph {et~al.}(2007)\citenamefont {Rini},
  \citenamefont {Tobey}, \citenamefont {Dean}, \citenamefont {Itatani},
  \citenamefont {Tomioka}, \citenamefont {Tokura}, \citenamefont {Schoenlein},\
  and\ \citenamefont {Cavalleri}}]{Rini2007}%
  \BibitemOpen
  \bibfield  {author} {\bibinfo {author} {\bibfnamefont {M.}~\bibnamefont
  {Rini}}, \bibinfo {author} {\bibfnamefont {R.}~\bibnamefont {Tobey}},
  \bibinfo {author} {\bibfnamefont {N.}~\bibnamefont {Dean}}, \bibinfo {author}
  {\bibfnamefont {J.}~\bibnamefont {Itatani}}, \bibinfo {author} {\bibfnamefont
  {Y.}~\bibnamefont {Tomioka}}, \bibinfo {author} {\bibfnamefont
  {Y.}~\bibnamefont {Tokura}}, \bibinfo {author} {\bibfnamefont {R.~W.}\
  \bibnamefont {Schoenlein}},\ and\ \bibinfo {author} {\bibfnamefont
  {A.}~\bibnamefont {Cavalleri}},\ }\bibfield  {title} {\bibinfo {title}
  {Control of the electronic phase of a manganite by mode-selective vibrational
  excitation},\ }\href {https://doi.org/10.1038/nature06119} {\bibfield
  {journal} {\bibinfo  {journal} {Nature}\ }\textbf {\bibinfo {volume} {449}},\
  \bibinfo {pages} {72} (\bibinfo {year} {2007})}\BibitemShut {NoStop}%
\bibitem [{\citenamefont {Disa}\ \emph {et~al.}(2023)\citenamefont {Disa},
  \citenamefont {Curtis}, \citenamefont {Fechner}, \citenamefont {Liu},
  \citenamefont {von Hoegen}, \citenamefont {F\"orst}, \citenamefont {Nova},
  \citenamefont {Narang}, \citenamefont {Maljuk}, \citenamefont {Boris},
  \citenamefont {Keimer},\ and\ \citenamefont
  {Cavalleri}}]{disa_photoinduced_2023}%
  \BibitemOpen
  \bibfield  {author} {\bibinfo {author} {\bibfnamefont {A.~S.}\ \bibnamefont
  {Disa}}, \bibinfo {author} {\bibfnamefont {J.}~\bibnamefont {Curtis}},
  \bibinfo {author} {\bibfnamefont {M.}~\bibnamefont {Fechner}}, \bibinfo
  {author} {\bibfnamefont {A.}~\bibnamefont {Liu}}, \bibinfo {author}
  {\bibfnamefont {A.}~\bibnamefont {von Hoegen}}, \bibinfo {author}
  {\bibfnamefont {M.}~\bibnamefont {F\"orst}}, \bibinfo {author} {\bibfnamefont
  {T.~F.}\ \bibnamefont {Nova}}, \bibinfo {author} {\bibfnamefont
  {P.}~\bibnamefont {Narang}}, \bibinfo {author} {\bibfnamefont
  {A.}~\bibnamefont {Maljuk}}, \bibinfo {author} {\bibfnamefont {A.~V.}\
  \bibnamefont {Boris}}, \bibinfo {author} {\bibfnamefont {B.}~\bibnamefont
  {Keimer}},\ and\ \bibinfo {author} {\bibfnamefont {A.}~\bibnamefont
  {Cavalleri}},\ }\bibfield  {title} {\bibinfo {title} {Photo-induced
  high-temperature ferromagnetism in {YTiO}$_3$},\ }\href
  {https://doi.org/10.1038/s41586-023-05853-8} {\bibfield  {journal} {\bibinfo
  {journal} {Nature}\ }\textbf {\bibinfo {volume} {617}},\ \bibinfo {pages}
  {73} (\bibinfo {year} {2023})}\BibitemShut {NoStop}%
\bibitem [{\citenamefont {Basini}\ \emph {et~al.}(2024)\citenamefont {Basini},
  \citenamefont {Pancaldi}, \citenamefont {Wehinger}, \citenamefont {Udina},
  \citenamefont {Unikandanunni}, \citenamefont {Tadano}, \citenamefont
  {Hoffmann}, \citenamefont {Balatsky},\ and\ \citenamefont
  {Bonetti}}]{Basini2024}%
  \BibitemOpen
  \bibfield  {author} {\bibinfo {author} {\bibfnamefont {M.}~\bibnamefont
  {Basini}}, \bibinfo {author} {\bibfnamefont {M.}~\bibnamefont {Pancaldi}},
  \bibinfo {author} {\bibfnamefont {B.}~\bibnamefont {Wehinger}}, \bibinfo
  {author} {\bibfnamefont {M.}~\bibnamefont {Udina}}, \bibinfo {author}
  {\bibfnamefont {V.}~\bibnamefont {Unikandanunni}}, \bibinfo {author}
  {\bibfnamefont {T.}~\bibnamefont {Tadano}}, \bibinfo {author} {\bibfnamefont
  {M.~C.}\ \bibnamefont {Hoffmann}}, \bibinfo {author} {\bibfnamefont {A.~V.}\
  \bibnamefont {Balatsky}},\ and\ \bibinfo {author} {\bibfnamefont
  {S.}~\bibnamefont {Bonetti}},\ }\bibfield  {title} {\bibinfo {title}
  {Terahertz electric-field-driven dynamical multiferroicity in
  {S}r{T}i{O}$_3$},\ }\href {https://doi.org/10.1038/s41586-024-07175-9}
  {\bibfield  {journal} {\bibinfo  {journal} {Nature}\ }\textbf {\bibinfo
  {volume} {628}},\ \bibinfo {pages} {534} (\bibinfo {year}
  {2024})}\BibitemShut {NoStop}%
\bibitem [{\citenamefont {Hase}\ \emph {et~al.}(1993)\citenamefont {Hase},
  \citenamefont {Terasaki},\ and\ \citenamefont
  {Uchinokura}}]{PhysRevLett.70.3651}%
  \BibitemOpen
  \bibfield  {author} {\bibinfo {author} {\bibfnamefont {M.}~\bibnamefont
  {Hase}}, \bibinfo {author} {\bibfnamefont {I.}~\bibnamefont {Terasaki}},\
  and\ \bibinfo {author} {\bibfnamefont {K.}~\bibnamefont {Uchinokura}},\
  }\bibfield  {title} {\bibinfo {title} {Observation of the spin-{P}eierls
  transition in linear {C}u$^{2+}$ (spin-1/2) chains in an inorganic compound
  {C}u{G}e{O}$_{3}$},\ }\href {https://doi.org/10.1103/PhysRevLett.70.3651}
  {\bibfield  {journal} {\bibinfo  {journal} {Phys. Rev. Lett.}\ }\textbf
  {\bibinfo {volume} {70}},\ \bibinfo {pages} {3651} (\bibinfo {year}
  {1993})}\BibitemShut {NoStop}%
\bibitem [{\citenamefont {Braden}\ \emph {et~al.}(1996)\citenamefont {Braden},
  \citenamefont {Wilkendorf}, \citenamefont {Lorenzana}, \citenamefont
  {A\"{\i}n}, \citenamefont {McIntyre}, \citenamefont {Behruzi}, \citenamefont
  {Heger}, \citenamefont {Dhalenne},\ and\ \citenamefont
  {Revcolevschi}}]{PhysRevB.54.1105}%
  \BibitemOpen
  \bibfield  {author} {\bibinfo {author} {\bibfnamefont {M.}~\bibnamefont
  {Braden}}, \bibinfo {author} {\bibfnamefont {G.}~\bibnamefont {Wilkendorf}},
  \bibinfo {author} {\bibfnamefont {J.}~\bibnamefont {Lorenzana}}, \bibinfo
  {author} {\bibfnamefont {M.}~\bibnamefont {A\"{\i}n}}, \bibinfo {author}
  {\bibfnamefont {G.~J.}\ \bibnamefont {McIntyre}}, \bibinfo {author}
  {\bibfnamefont {M.}~\bibnamefont {Behruzi}}, \bibinfo {author} {\bibfnamefont
  {G.}~\bibnamefont {Heger}}, \bibinfo {author} {\bibfnamefont
  {G.}~\bibnamefont {Dhalenne}},\ and\ \bibinfo {author} {\bibfnamefont
  {A.}~\bibnamefont {Revcolevschi}},\ }\bibfield  {title} {\bibinfo {title}
  {Structural analysis of {C}u{G}e{O}$_{3}$: Relation between nuclear structure
  and magnetic interaction},\ }\href {https://doi.org/10.1103/PhysRevB.54.1105}
  {\bibfield  {journal} {\bibinfo  {journal} {Phys. Rev. B}\ }\textbf {\bibinfo
  {volume} {54}},\ \bibinfo {pages} {1105} (\bibinfo {year}
  {1996})}\BibitemShut {NoStop}%
\bibitem [{\citenamefont {Kraus}\ \emph {et~al.}(2010)\citenamefont {Kraus},
  \citenamefont {B\"uchner}, \citenamefont {Knupfer}, \citenamefont {Glawion},
  \citenamefont {Sing},\ and\ \citenamefont {Claessen}}]{PhysRevB.81.125133}%
  \BibitemOpen
  \bibfield  {author} {\bibinfo {author} {\bibfnamefont {R.}~\bibnamefont
  {Kraus}}, \bibinfo {author} {\bibfnamefont {B.}~\bibnamefont {B\"uchner}},
  \bibinfo {author} {\bibfnamefont {M.}~\bibnamefont {Knupfer}}, \bibinfo
  {author} {\bibfnamefont {S.}~\bibnamefont {Glawion}}, \bibinfo {author}
  {\bibfnamefont {M.}~\bibnamefont {Sing}},\ and\ \bibinfo {author}
  {\bibfnamefont {R.}~\bibnamefont {Claessen}},\ }\bibfield  {title} {\bibinfo
  {title} {Anisotropic crystal field, {M}ott gap, and interband excitations in
  {T}i{OC}l: An electron energy-loss study},\ }\href
  {https://doi.org/10.1103/PhysRevB.81.125133} {\bibfield  {journal} {\bibinfo
  {journal} {Phys. Rev. B}\ }\textbf {\bibinfo {volume} {81}},\ \bibinfo
  {pages} {125133} (\bibinfo {year} {2010})}\BibitemShut {NoStop}%
\bibitem [{\citenamefont {Sachdev}\ and\ \citenamefont
  {Bhatt}(1990)}]{sachdev_bond-operator_1990}%
  \BibitemOpen
  \bibfield  {author} {\bibinfo {author} {\bibfnamefont {S.}~\bibnamefont
  {Sachdev}}\ and\ \bibinfo {author} {\bibfnamefont {R.~N.}\ \bibnamefont
  {Bhatt}},\ }\bibfield  {title} {\bibinfo {title} {Bond-operator
  representation of quantum spins: {M}ean-field theory of frustrated quantum
  {H}eisenberg antiferromagnets},\ }\href
  {https://doi.org/10.1103/PhysRevB.41.9323} {\bibfield  {journal} {\bibinfo
  {journal} {Phys. Rev. B}\ }\textbf {\bibinfo {volume} {41}},\ \bibinfo
  {pages} {9323} (\bibinfo {year} {1990})}\BibitemShut {NoStop}%
\bibitem [{\citenamefont {Gopalan}\ \emph {et~al.}(1994)\citenamefont
  {Gopalan}, \citenamefont {Rice},\ and\ \citenamefont
  {Sigrist}}]{gopalan_spin_1994}%
  \BibitemOpen
  \bibfield  {author} {\bibinfo {author} {\bibfnamefont {S.}~\bibnamefont
  {Gopalan}}, \bibinfo {author} {\bibfnamefont {T.~M.}\ \bibnamefont {Rice}},\
  and\ \bibinfo {author} {\bibfnamefont {M.}~\bibnamefont {Sigrist}},\
  }\bibfield  {title} {\bibinfo {title} {Spin ladders with spin gaps: A
  description of a class of cuprates},\ }\href
  {https://doi.org/10.1103/PhysRevB.49.8901} {\bibfield  {journal} {\bibinfo
  {journal} {Phys. Rev. B}\ }\textbf {\bibinfo {volume} {49}},\ \bibinfo
  {pages} {8901} (\bibinfo {year} {1994})}\BibitemShut {NoStop}%
\bibitem [{\citenamefont {Kumar}(2010)}]{kumar_bond_2010}%
  \BibitemOpen
  \bibfield  {author} {\bibinfo {author} {\bibfnamefont {B.}~\bibnamefont
  {Kumar}},\ }\bibfield  {title} {\bibinfo {title} {Bond operators and triplon
  analysis for spin-${S}$ dimer antiferromagnets},\ }\href
  {https://doi.org/10.1103/PhysRevB.82.054404} {\bibfield  {journal} {\bibinfo
  {journal} {Phys. Rev. B}\ }\textbf {\bibinfo {volume} {82}},\ \bibinfo
  {pages} {054404} (\bibinfo {year} {2010})}\BibitemShut {NoStop}%
\bibitem [{\citenamefont {Kubo}(1962)}]{kubo_generalized_1962}%
  \BibitemOpen
  \bibfield  {author} {\bibinfo {author} {\bibfnamefont {R.}~\bibnamefont
  {Kubo}},\ }\bibfield  {title} {\bibinfo {title} {Generalized {Cumulant}
  {Expansion} {Method}},\ }\href {https://doi.org/10.1143/JPSJ.17.1100}
  {\bibfield  {journal} {\bibinfo  {journal} {Journal of the Physical Society
  of Japan}\ }\textbf {\bibinfo {volume} {17}},\ \bibinfo {pages} {1100}
  (\bibinfo {year} {1962})}\BibitemShut {NoStop}%
\bibitem [{\citenamefont {Thompson}\ and\ \citenamefont
  {Phillips}(2009)}]{thompson2009dynamo}%
  \BibitemOpen
  \bibfield  {author} {\bibinfo {author} {\bibfnamefont {S.}~\bibnamefont
  {Thompson}}\ and\ \bibinfo {author} {\bibfnamefont {T.}~\bibnamefont
  {Phillips}},\ }\href {https://books.google.com/books?id=dKVbT-ZmdDwC} {\emph
  {\bibinfo {title} {Dynamo-Electric Machinery; A Manual for Students of
  Electrotechnics}}}\ (\bibinfo  {publisher} {BiblioBazaar},\ \bibinfo {year}
  {2009})\BibitemShut {NoStop}%
\bibitem [{\citenamefont {Dolgner}\ \emph {et~al.}(2026)\citenamefont
  {Dolgner}, \citenamefont {Manske}, \citenamefont {Freericks},\ and\
  \citenamefont {Yarmohammadi}}]{Zenodo}%
  \BibitemOpen
  \bibfield  {author} {\bibinfo {author} {\bibfnamefont {J.}~\bibnamefont
  {Dolgner}}, \bibinfo {author} {\bibfnamefont {D.}~\bibnamefont {Manske}},
  \bibinfo {author} {\bibfnamefont {J.~K.}\ \bibnamefont {Freericks}},\ and\
  \bibinfo {author} {\bibfnamefont {M.}~\bibnamefont {Yarmohammadi}},\
  }\bibfield  {title} {\bibinfo {title} {[{D}ata set]},\ }\bibfield  {journal}
  {\bibinfo  {journal} {Zenodo}\ }\href
  {https://doi.org/10.5281/zenodo.22069073} {10.5281/zenodo.22069073} (\bibinfo
  {year} {2026})\BibitemShut {NoStop}%
\bibitem [{\citenamefont {{Van Kampen}}(1974)}]{van_kampen_cumulant_1974}%
  \BibitemOpen
  \bibfield  {author} {\bibinfo {author} {\bibfnamefont {N.}~\bibnamefont {{Van
  Kampen}}},\ }\bibfield  {title} {\bibinfo {title} {A cumulant expansion for
  stochastic linear differential equations. i},\ }\href
  {https://doi.org/https://doi.org/10.1016/0031-8914(74)90121-9} {\bibfield
  {journal} {\bibinfo  {journal} {Physica}\ }\textbf {\bibinfo {volume} {74}},\
  \bibinfo {pages} {215} (\bibinfo {year} {1974})}\BibitemShut {NoStop}%
\bibitem [{\citenamefont {Kerber}\ \emph {et~al.}(2025)\citenamefont {Kerber},
  \citenamefont {Ritsch},\ and\ \citenamefont
  {Ostermann}}]{kerber2025cumulantsexpansionapproachgood}%
  \BibitemOpen
  \bibfield  {author} {\bibinfo {author} {\bibfnamefont {J.}~\bibnamefont
  {Kerber}}, \bibinfo {author} {\bibfnamefont {H.}~\bibnamefont {Ritsch}},\
  and\ \bibinfo {author} {\bibfnamefont {L.}~\bibnamefont {Ostermann}},\
  }\bibfield  {title} {\bibinfo {title} {The cumulants expansion approach: The
  good, the bad and the ugly},\ }\href {https://arxiv.org/abs/2511.20115} {\
  (\bibinfo {year} {2025})},\ \Eprint {https://arxiv.org/abs/2511.20115}
  {arXiv:2511.20115} \BibitemShut {NoStop}%
\end{thebibliography}%

\end{document}